\documentclass[12pt]{iopart}
\usepackage{setspace}

\usepackage{dcolumn}   
\usepackage[T1]{fontenc}
\usepackage[explicit]{titlesec}
\usepackage{xcolor}
\usepackage{textcase}
\usepackage{lmodern}
\usepackage{lipsum}
\usepackage[caption=false,labelformat=simple]{subfig}

\usepackage{graphicx}  
\usepackage{amsmath}

\usepackage{amssymb}   

\usepackage{slashed}
\usepackage{stmaryrd}
\usepackage{color}
\usepackage[ugly]{units}
\usepackage{nicefrac}
\usepackage{lipsum}
\usepackage[utf8]{inputenc}

\providecommand{\abs}[1]{\left\vert#1\right\vert}

\newcommand{\new}[1]{{#1}}
\newcommand{\neww}[1]{{#1}}

\newcommand{\mychar}[1]{%
  \begingroup\normalfont
  \includegraphics[height=\fontcharht\font`\B]{#1}%
  \endgroup
}
\newcommand{\mycharp}[1]{%
  \begingroup\normalfont
  \includegraphics[height=0.5\fontcharht\font`\B]{#1}%
  \endgroup
}
\newcommand{\mycharv}[1]{%
  \begingroup\normalfont
  \includegraphics[height=1.5\fontcharht\font`\B ]{#1}%
  \endgroup
}
\begin{document}
\title{Laser pulse-length effects in trident pair production}
\author{U. Hernandez Acosta and B. Kämpfer}
\author{}
\address{\mbox{Helmholtz-Zentrum Dresden-Rossendorf, POB\,51\,01\,19, 01314 Dresden, Germany}\\
TU Dresden, Institut für Theoretische Physik, 01062 Dresden, Germany}
\ead{u.hernandez@hzdr.de}
\vspace{10pt}
\begin{indented}
\item[]\today
\end{indented}
\begin{abstract}
 Laser pulses facilitate multiphoton contributions to the trident pair production \( e_L^- \to e_L^- + e_L^+ + e_L^- \), where the label \( L \) indicates a laser field dressed electron (\( e^-\)) or positron (\( e^+ \)). We isolate the impact of the pulse envelope in the trident \( S \) matrix element, formulated within the Furry picture, in leading order of a series expansion in the classical non-linearity parameter \( a_0 \). Generally, the Fourier transform of the envelope carries the information on the pulse length, which becomes an easily tractable function in the case of a \( \cos^2 \) pulse envelope. The transition to a monochromatic laser wave can be handled in a transparent manner, as also the onset of \neww{bandwidth} effects for short pulses can be factorized out and studied separately.
\end{abstract}

\maketitle

%
%

\section{Introduction}
High-intensity laser beams in the optical regime are customarily generated by the chirped pulse amplification (cf. \cite{Mourou:2006}). Intensities up to \( \unitfrac[10^{22}]{W}{cm^2}\) are achievable nowadays in several laboratories \cite{Yanovsky:2008}, yielding a classical non-linearity parameter of \( a_0 = \mathcal O(10-50) \) in the focal spot\footnote{\new{The relation of \( a_0 \) vs. the laser peak intensity \( I_L \) and frequency \( \omega \) reads \( a_0\simeq 7.5 \frac{\text{eV}}{ \omega}\sqrt{\frac{I_L}{10^{20}\text{W}/\text{cm}^2}}\), cf.\ \cite{Di-Piazza:2012}.}}. Ongoing projects \cite{ELI:2011,Papadopoulos:2016,Yu:2018} of \( \unit[10]{PW} \) class lasers envisage even larger values of \( a_0 \). Due to higher frequencies in XFEL beams, \( \omega= \mathcal O(\unit[10]{keV}) \), the parameter \( a_0 \) stays significantly below unity, despite similar intensities of \( \mathcal O(\unitfrac[10^{22}]{W}{cm^2}) \) \new{when tight focusing is attained \cite{Ringwald:2001}}. Given such a variety of laser facilities, the experimental exploration of nonlinear QED effects became feasible and is currently further promoted. Elementary processes are under consideration with the goal of testing QED in the strong-field regime. Most notable is the nonlinear Compton process \( e_L^- \to e_L^- + \gamma \), also w.r.t.\ the subsequent use of the high energy photons (\( \gamma \)), up to prospects of industrial applications. While in the pioneering theoretical studies \cite{Nikishov:1964,Ritus:1985} the higher harmonics, related to multi-photon effects, i.e. the simultaneous interaction of the electron with a multitude of photons, in monochromatic laser beams have been considered, the study of laser pulses revealed a multitude of novel structures in the \( \gamma \) spectrum \cite{Di-Piazza:2012,Mackenroth:2011, Krajewska:2014, Seipt:2016,Titov:2016, Seipt:2014, Seipt:2013,Seipt:2011,Heinzl:2010,Narozhnyi:1974}. \\The nonlinear Breit-Wheeler process, \( \gamma\to e_L^- + e_L^+ \) \cite{Reiss:1962,Reiss:1971,Di-Piazza:2012,Titov:2016r,Titov:2016,Titov:2013,Titov:2012,Heinzl:2010b,Krajewska:2012, Kaminski:2018,Jansen:2016,Jansen:2017,Jansen:2016b}, as cross channel of the nonlinear Compton process, is in contrast a threshold process -- sometimes termed a genuine quantum process -- since the probe photon \( \gamma \) energy in combination with the laser must supply sufficient energy to produce a \( e^+e^- \) pair. When considering the seminal SLAC experiment E-144 \cite{Bamber:1999,Burke:1997} as a two-step process (first step: generation of a high-energy photon \( \gamma \) by Compton backscattering \cite{Bula:1996}, second step: Breit-Wheeler process \( \gamma+ L \to e^+e^-\)), also the nonlinear Breit-Wheeler process has been identified with the simultaneous interaction of up to five photons in the elementary subprocess.\\
Strictly speaking, the mentioned two-step process is only a part of trident pair production \( e_L^-\to e_L^- + e_L^+ + e_L^- \), as stressed in \cite{Hu:2010, Ilderton:2011}. Since the trident process is the starting point of seeded QED avalanches, expected to set in at high-intensities, it is currently a subject of throughout analyses \cite{Dinu:2018,King:2018,Mackenroth:2018}, also for benchmarking PIC codes \cite{Blackburn:2017}.\\\new{Given the high repetition rate of the European XFEL \cite{XFEL:2007} a potentially interesting option is to combine it with a synchronized electron beam of about \( \unit[50]{MeV} \) (to operate slightly above the threshold) in order to facilitate a high-statistics search for the dark photon. Such a dark photon (also dubbed \( U \) boson or hidden photon) is a candidate for Dark Matter beyond the standard model of particle physics; it is a possible extension which enjoys intense theoretical \cite{Rizzo:2018,Bauer:2018,Denig:2016} and experimental \cite{Beranek:2013,Curciarello:2016,Adrian:2018,Raggi:2018} considerations. A corresponding theoretical analysis of the trident process can be found in \cite{Gakh:2018}. In fact, the trident process -- in a perturbative QED language -- includes sub-diagrams of the type \( \gamma^*\to e^+e^- \), i.e.\ an intermediate (virtual) photon which decays into a \( e^+e^- \) pair. Via kinetic mixing, that virtual photon may ``temporarily'' couple to a dark photon \( A' \), e.g. \( \gamma^*\to A'\to \gamma^* \), thus signalizing its presence as a peak of the invariant mass distribution of \( e^+e^- \). The peak would be at the mass of the dark photon and its  width is related to the kinetic mixing strength.}\\
We briefly mention the trident option of the LUXE project \cite{Wing:2017,Hartin:2019} at DESY/Hamburg, which however is primarily dedicated to explore the ``boiling of the vacuum'' by means of the nonlinear Breit-Wheeler process in the Ritus corner, i.e.\ a kinematical region with a nonperturbative field strength dependence and coupling constant \( \abs{ e} \) dependence analog to the Schwinger pair creation probability.\\
While most of the above quoted papers focus on nonlinear effects in strong laser pulses, that is the impact of multiphoton contributions, we aim here at the study of \new{apparent} multiphoton effects \new{due to bandwidth effects} in \new{weak and} moderately strong laser pulses with \( a_0<1 \). The analysis of the Breit-Wheeler pair production in \cite{Titov:2013,Nousch:2012} revealed that in such a regime interesting features appear for short and ultra-short pulses. For instance, despite of \( a_0<1 \) a significant subthreshold pair production is enabled. Roughly speaking, in short pulses the frequency spectrum contains high Fourier components, thus enabling the subthreshold pair creation.\ In that respect, we are going to study the relevance of the pulse duration for the trident process.\ In contrast to the elementary one-vertex processes, the trident process as a two-vertex process obeys a higher complexity, similar to the two-photon Compton scattering \cite{Lotstedt:2009,Seipt:2012,Mackenroth:2013}.\\Our paper is organised as follows. In section \ref{sec:matEl}, the matrix element is evaluated with emphasis on a certain regularisation required to uncover the perturbative limit. The weak-field expansion is presented in section \ref{sec:wf}, where the Fourier transform of the background field amplitude is highlighted as a central quantity. The case of a \( \cos^2 \) envelope is elaborated in dome detail in section \ref{sec:cos2}, where also numerical examples are exhibited. The conclusion can be found in section \ref{sec:conc}.

\section{Matrix element in the Furry picture}\label{sec:matEl}
\begin{figure} [htb]
\begin{minipage}{0.35\textwidth}
\subfloat[]{\includegraphics[width=\textwidth]{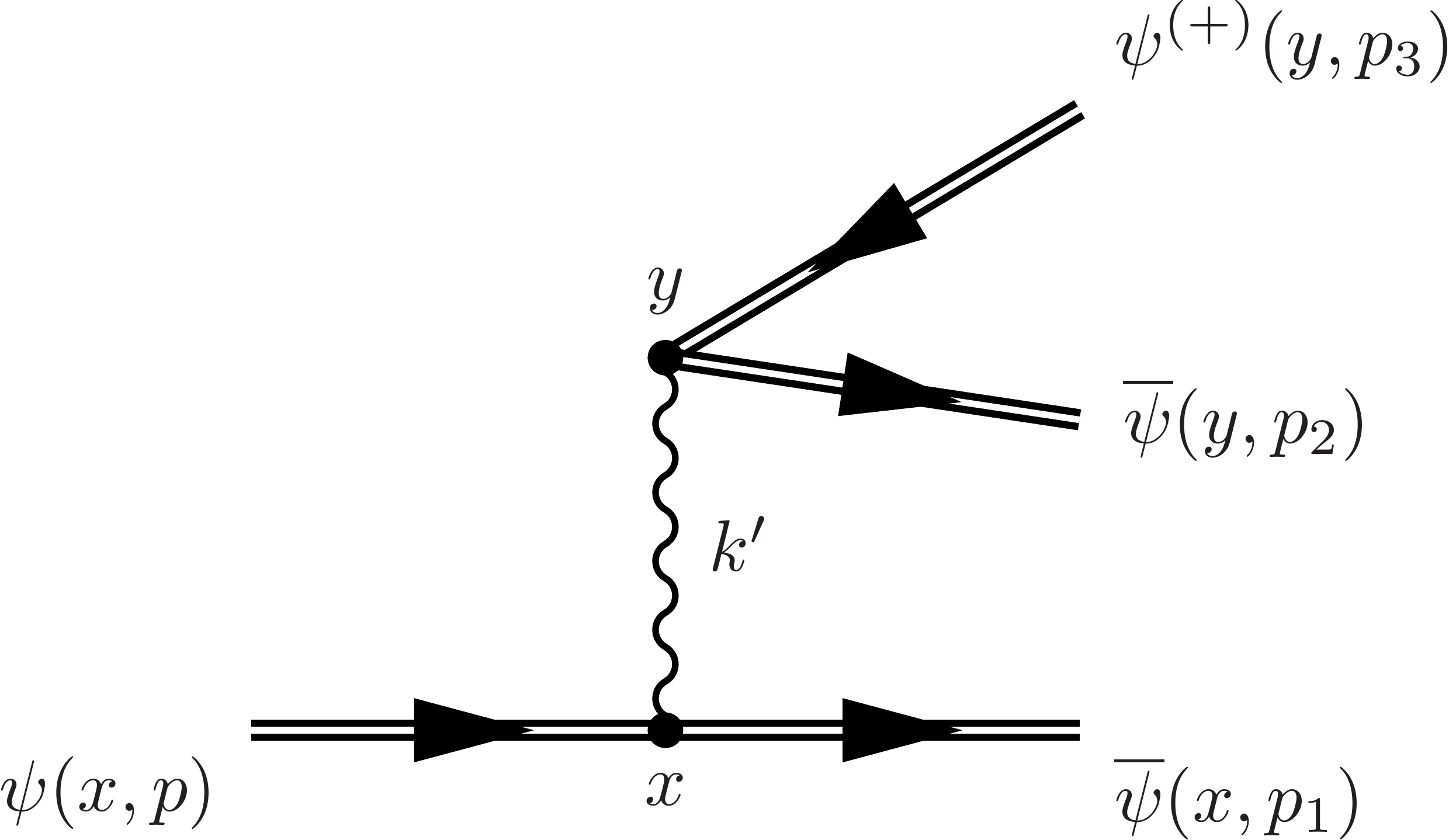}\label{fig:feynDiagFull-a}}
\end{minipage}
\( \Longrightarrow \)
\subfloat[]{\label{fig:feynDiagFull-b}
\begin{minipage}{0.28\textwidth}
\includegraphics[width=\textwidth]{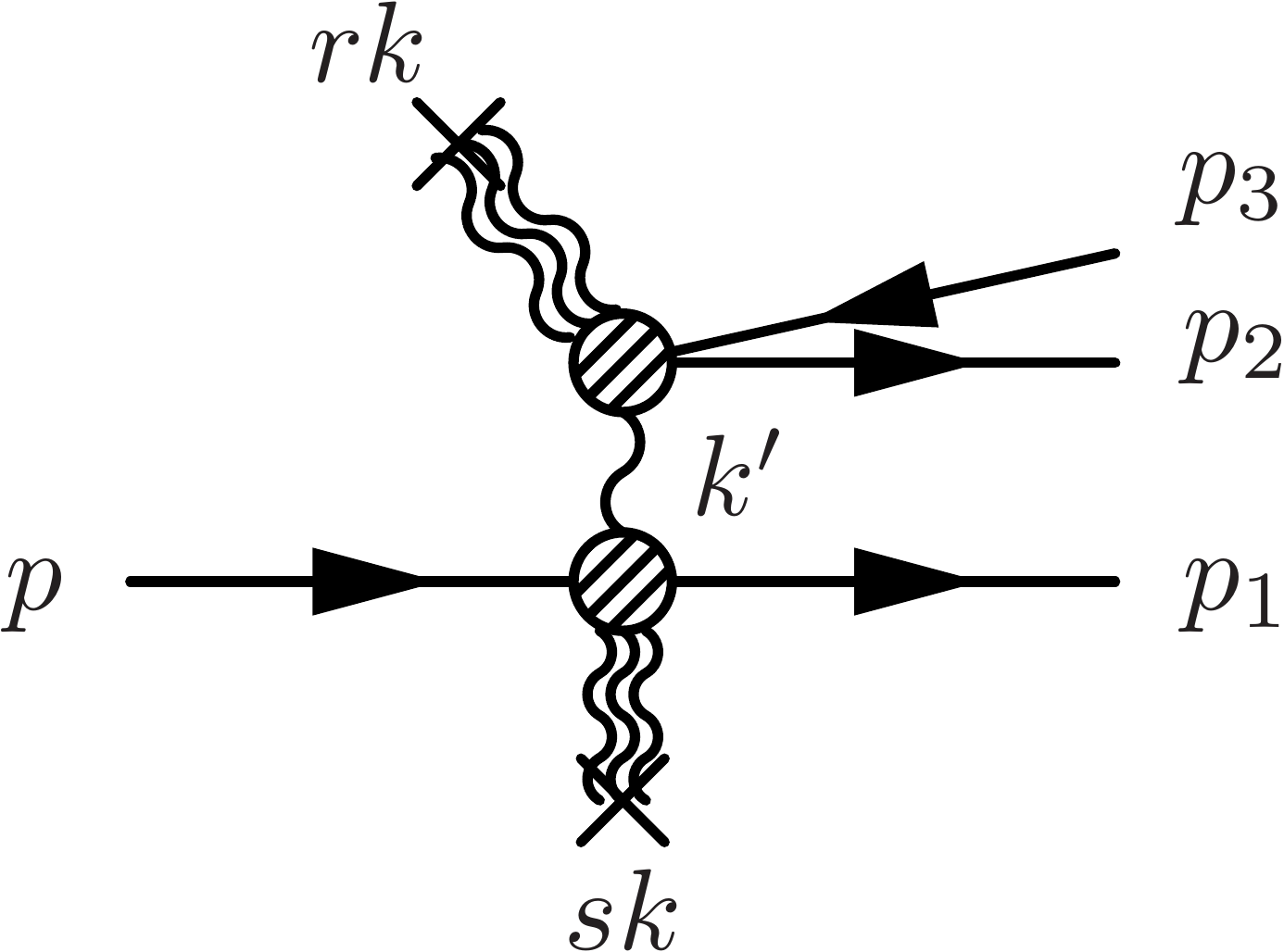}
\end{minipage}
\( \Longrightarrow \)
\begin{minipage}{0.28\textwidth}
\vspace{-5mm}
\includegraphics[width=\textwidth]{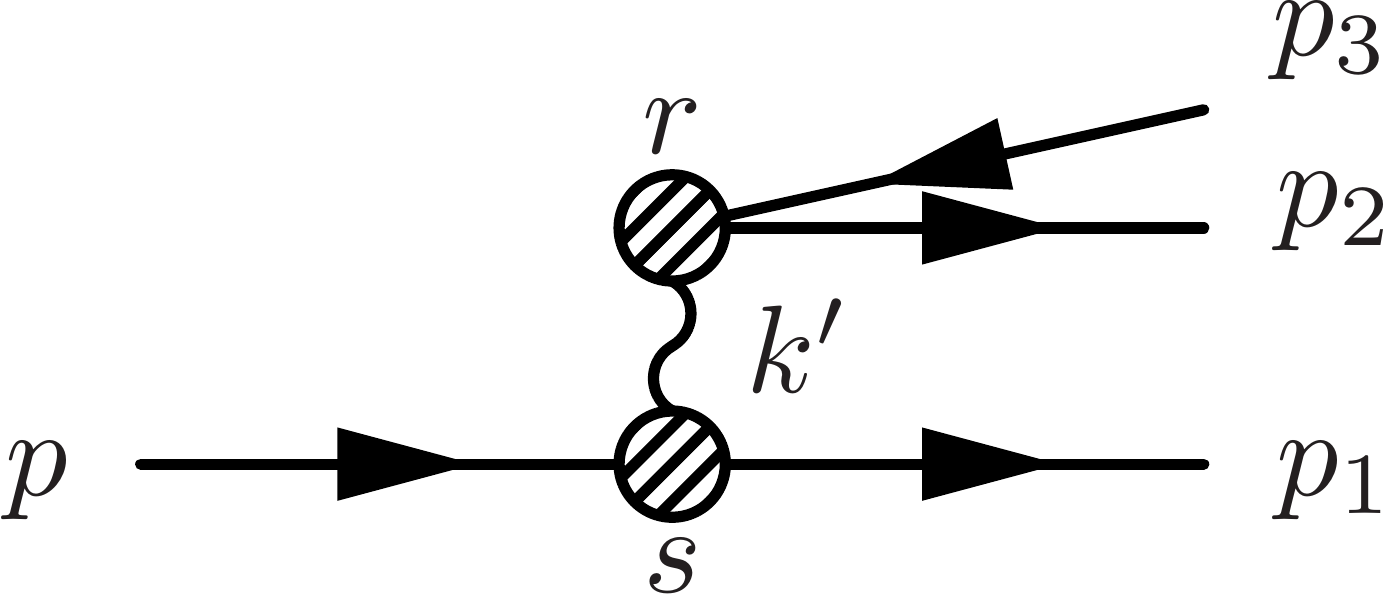}
\end{minipage}
}\\\\
=\subfloat[]{\label{fig:feynDiagFull-c}
\begin{minipage}{0.22\textwidth}
\includegraphics[width=\textwidth]{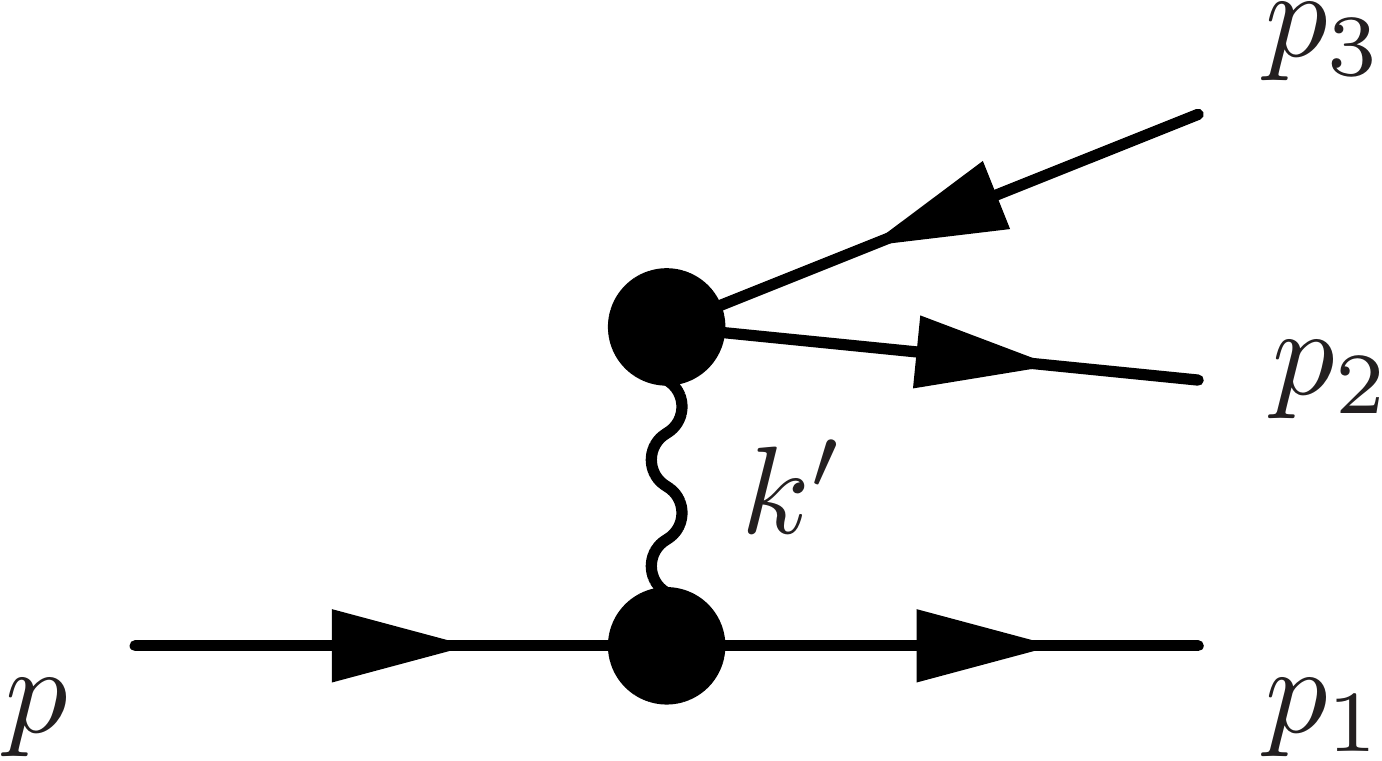}
\end{minipage}
\ +
\begin{minipage}{0.22\textwidth}
\includegraphics[width=\textwidth]{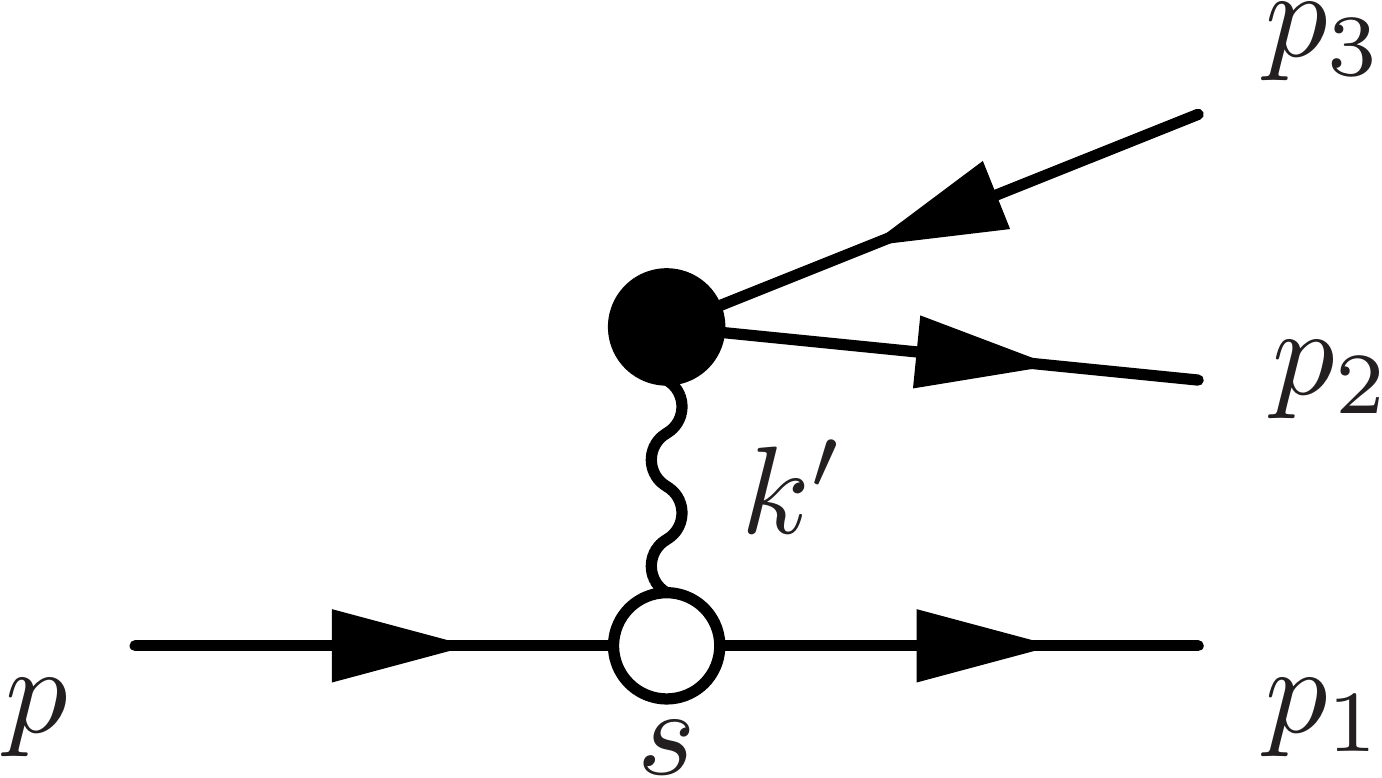}
\end{minipage}
\ +
\begin{minipage}{0.22\textwidth}
\includegraphics[width=\textwidth]{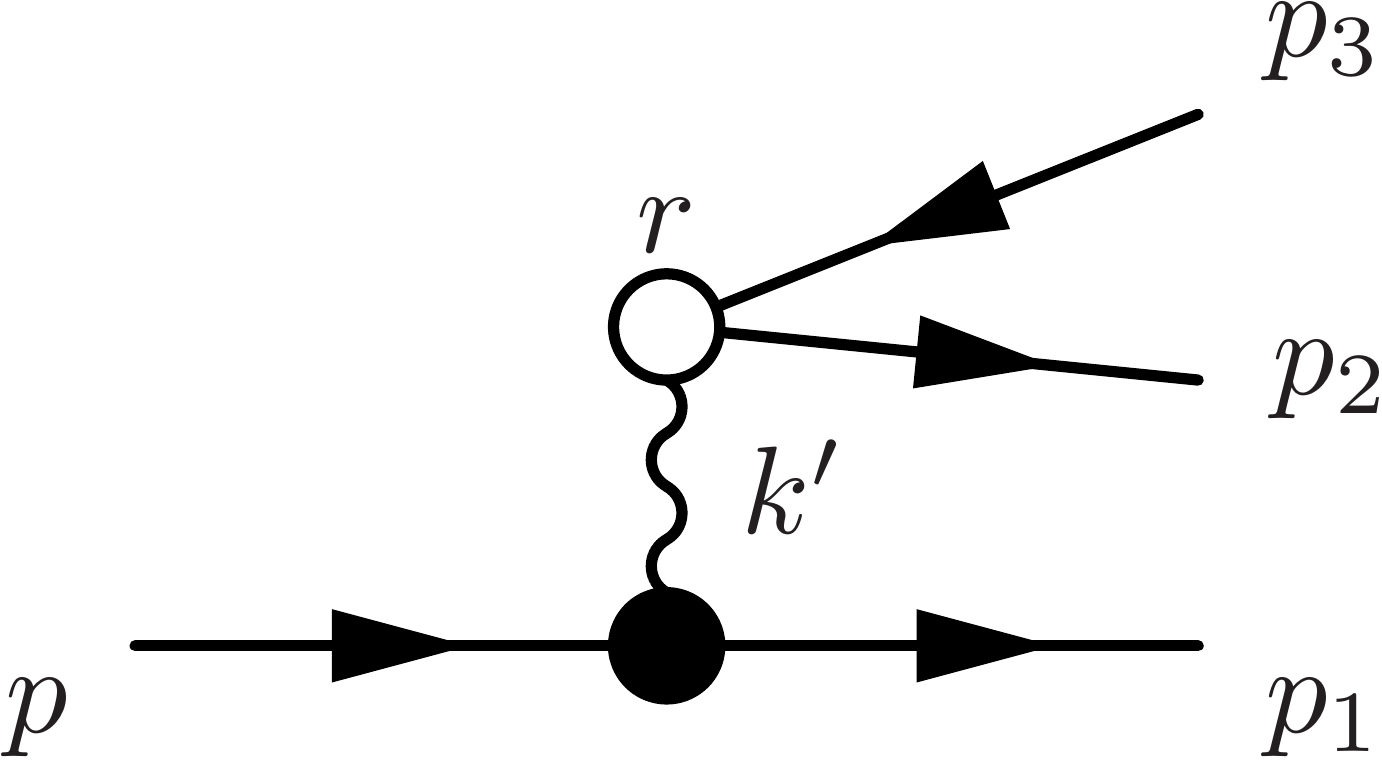}
\end{minipage}
\ +
\begin{minipage}{0.22\textwidth}
\includegraphics[width=\textwidth]{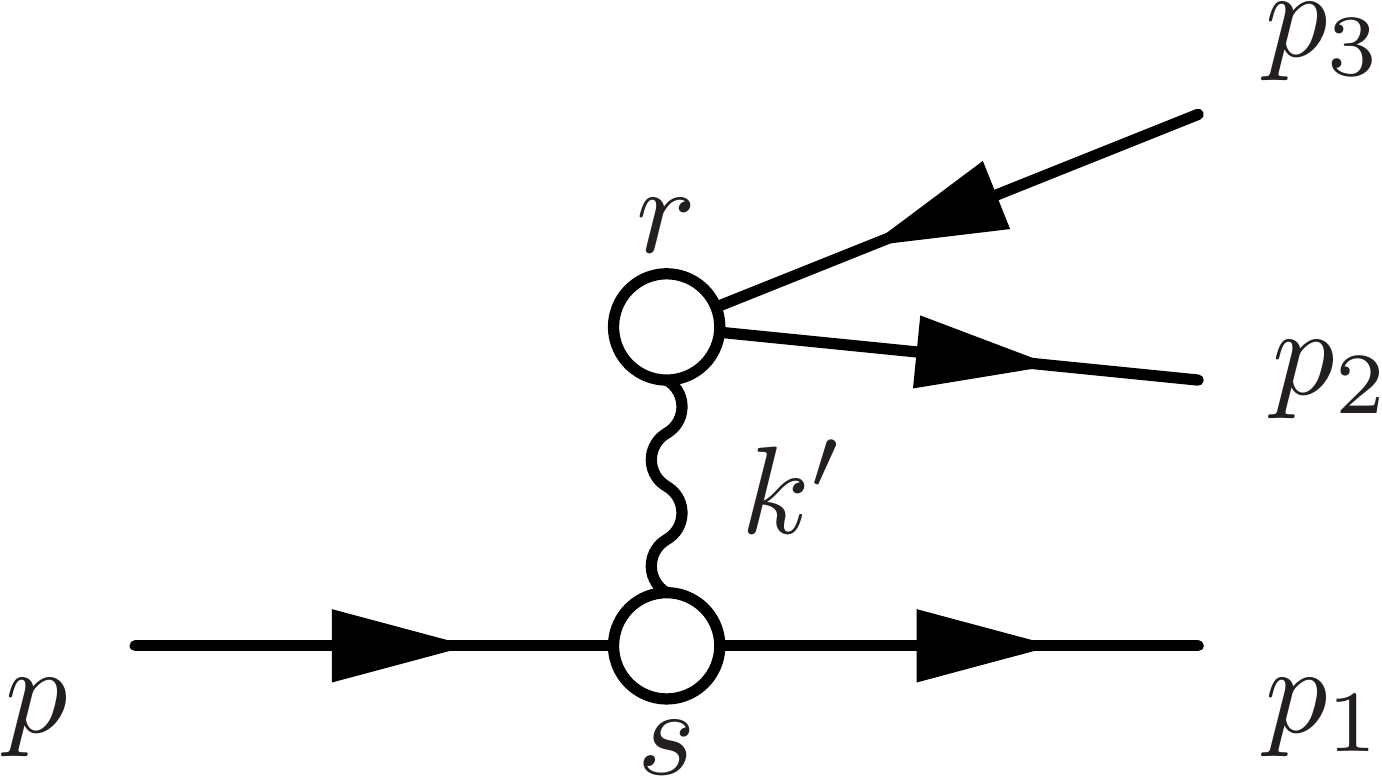}
\end{minipage}
}
\caption{Diagrams for the trident process. \subref*{fig:feynDiagFull-a} Lowest order Feynman diagram in position space Furry picture with \(\mychar{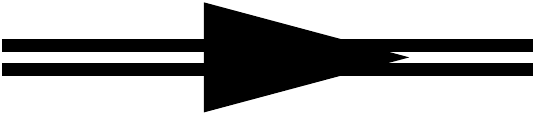}= \psi(x,p) \) for the Volkov state, \( \mycharp{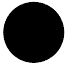} = -i e\int d^4x\, \gamma^\mu \) for the vertex at \( x \), \( \mycharp{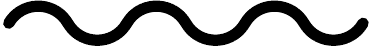} = D  _{ \mu \nu} (x-y) \) for the bare photon propagator. \subref*{fig:feynDiagFull-b} The \new{translation into} momentum space \new{to arrive at the right diagram} with \( \mychar{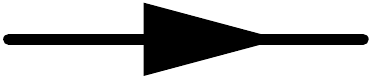} \) for the free Dirac spinor, \(  \mycharv{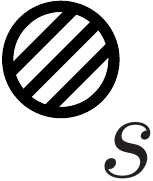} = -ie\int \frac{ds}{2 \pi} \Gamma ^\mu (s,p,p_1) \) with local four-momentum balance, e.g.\ \( p + sk = k' + p_1 \) and \(  \mycharp{bareProp} = D  _{ \mu \nu} (k') = \frac{i g  _{ \mu \nu} }{k^{'2} + i \epsilon} \) for the bare photon propagator in Feynman gauge. \subref*{fig:feynDiagFull-c} Using the regularized vertex \(  \mycharv{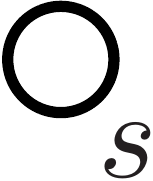} =  -ie\int \frac{ds}{2 \pi} \Gamma ^\mu_{\mathrm{reg}} (s,p,p_1)\) from \( \Gamma_0 ^\mu = \mathcal G \pi \delta(s) +\overset{12}{\Gamma ^\mu}_0  \) in \( \Gamma ^\mu = \Gamma ^\mu _0 +\Gamma ^\mu _1 + \Gamma ^\mu _2  \), i.e.\ \( \Gamma ^\mu _{ \mathrm{reg}} =\overset{12}{\Gamma ^\mu}_0  + \Gamma ^\mu _1 + \Gamma ^\mu _2 \) and the one photon vertex \( \mychar{qedVertex} =  \mathcal G \pi\, \delta(s)\gamma^\mu \). We note \( \Gamma^\mu_1 \propto ea_0, \Gamma ^\mu _2 \propto \left( ea_0\right)^2  \), while \( \overset{12}{\Gamma ^\mu}_0  \) has terms \( \propto ea_0 \) and \( \propto\left( ea_0\right)^2 \).}
\label{fig:feynDiagFull}
\end{figure} 
\begin{figure}[htb]
\begin{align*}
&\begin{minipage}{0.22\textwidth}
\includegraphics[width=\textwidth]{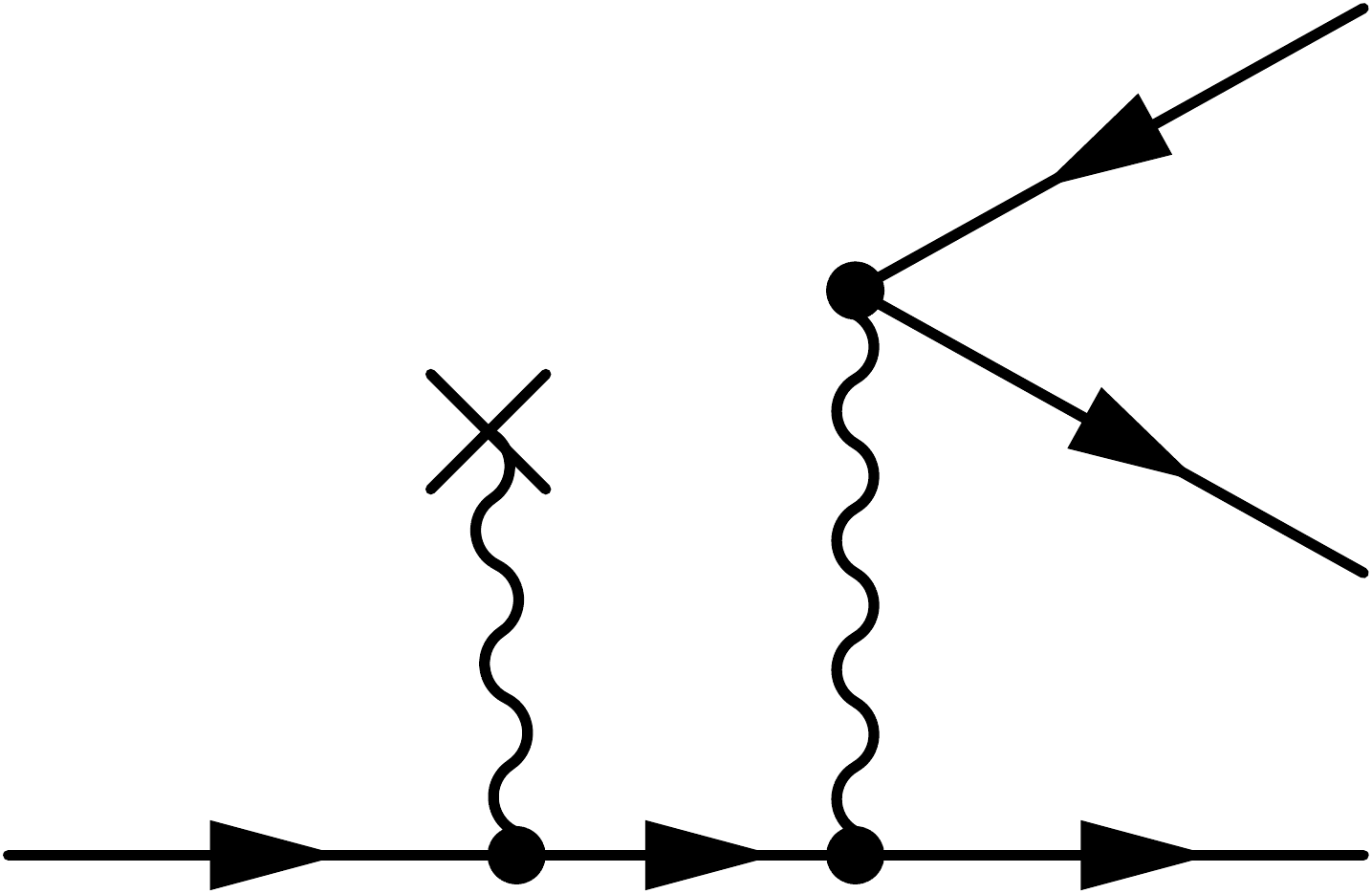}
\end{minipage}
\ +
\begin{minipage}{0.22\textwidth}
\includegraphics[width=\textwidth]{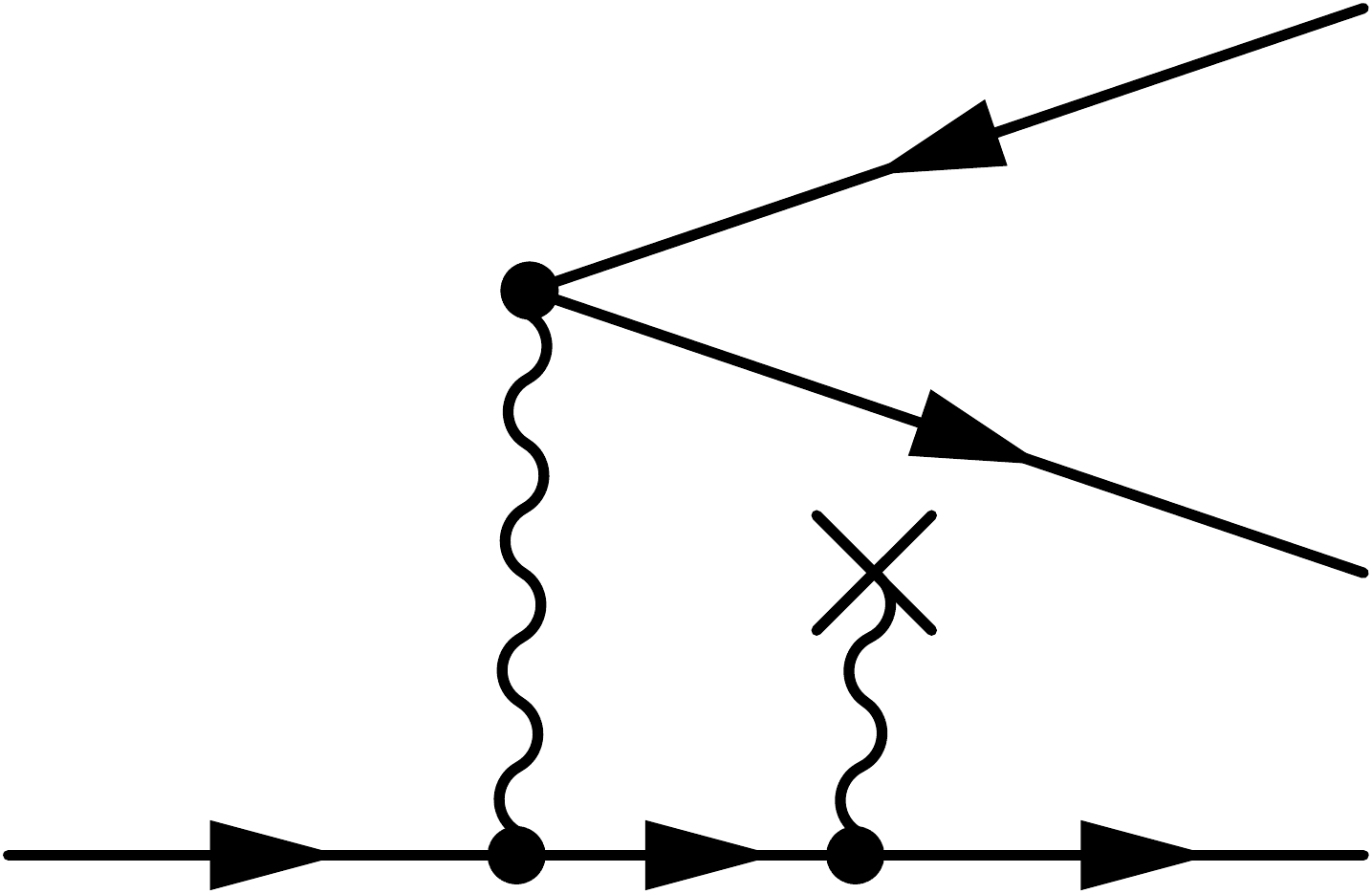}
\end{minipage}
\\\\ +
&\begin{minipage}{0.22\textwidth}
\includegraphics[width=\textwidth]{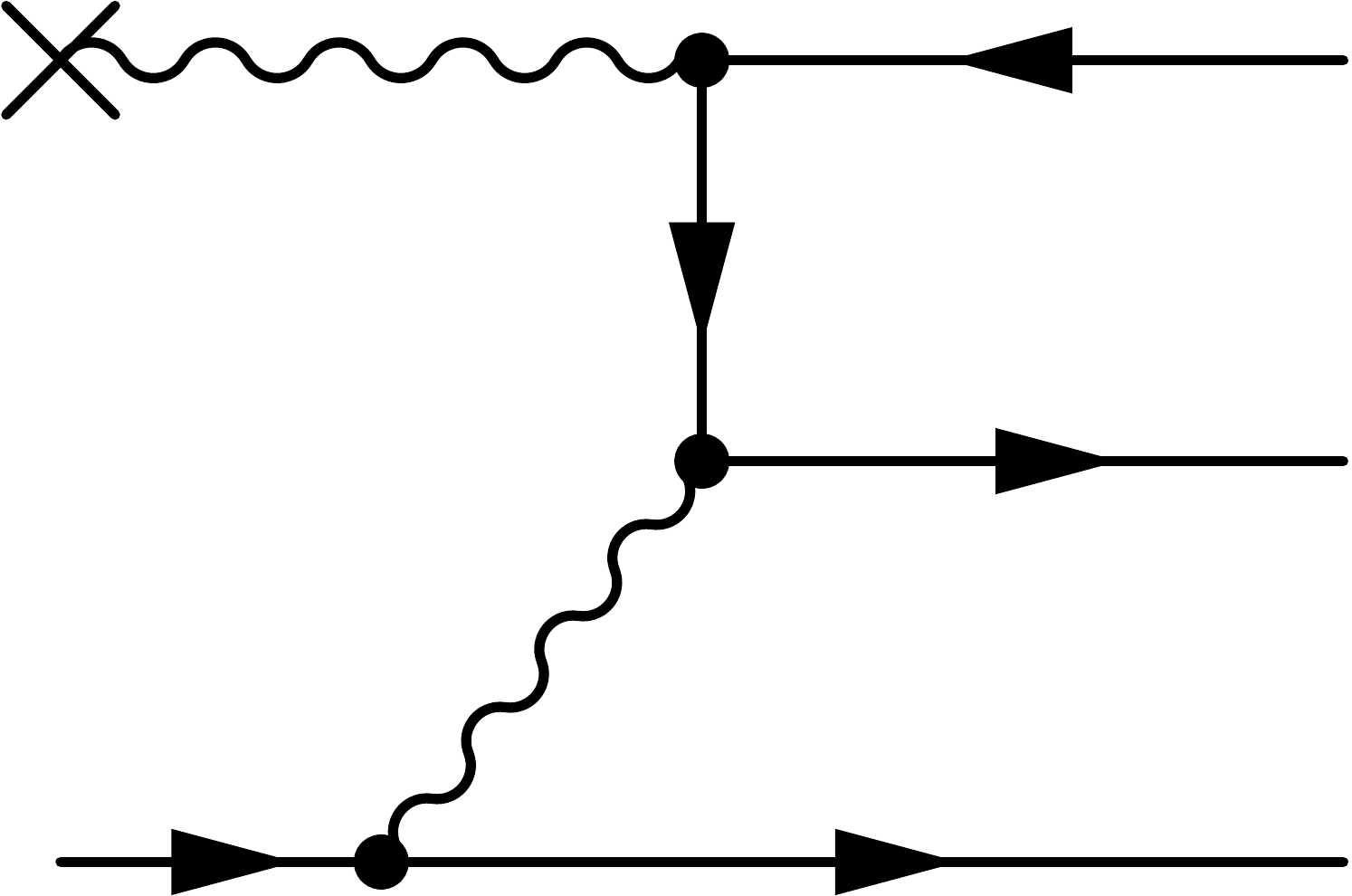}
\end{minipage}
\ +
\begin{minipage}{0.22\textwidth}
\includegraphics[width=\textwidth]{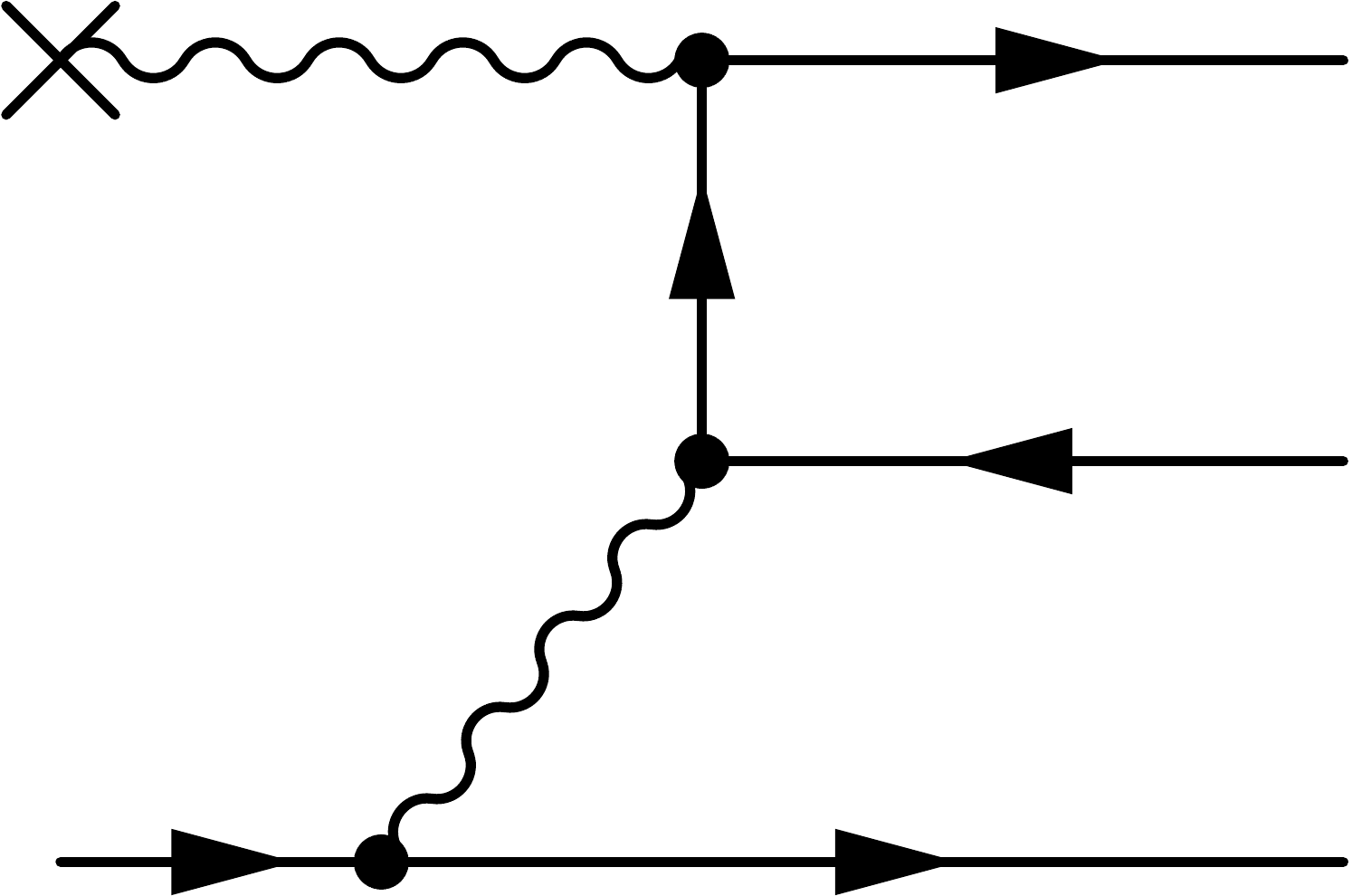}
\end{minipage}
\end{align*}
\caption{Leading order contributions in \( a_0 \) emerging from the second and third diagrams in figure \ref{fig:feynDiagFull-c} \new{when ignoring the pulse shape function}. They can be more directly generated by utilising the first-order iterative solution of the Lippmann-Schwinger equation for the Volkov solution.\ \new{T}hese diagrams \new{are} the standard perturbative QED Feynman diagrams. The fourth diagram in figure \ref{fig:feynDiagFull-c} becomes accordingly in leading order the set of here displayed diagrams, however with an additional external photon line  \( \mychar{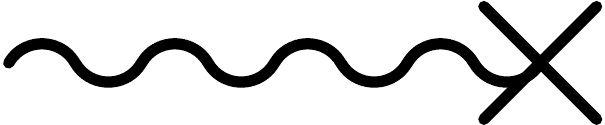} \)  attached to spinor lines in all possible combinations, thus making its contribution of order \( a_0^2 \) at least. As in figure \ref{fig:feynDiagFull} the electron antisymmetrizing contributions upon \( p_1\leftrightarrow p_2 \) are not exhibited.}
\label{fig:FeynDiagPert}
\end{figure}
The leading-order tree level Feynman diagram of the trident process is exhibited in figure \ref{fig:feynDiagFull-a}. The corresponding \( S \) matrix reads
\begin{align}\label{eq:SmatrixGeneral}
 S_{fi} = e^2 \int d^4x\int d^4y\,&\bigg[ \left(\overline \psi(x;p_1) \gamma^\mu \psi(x;p)\right) D_{ \mu \nu}(x-y) \left( \overline \psi(y;p_2) \gamma^\nu \psi^{(+)}(y;p_3) \right)\\&-(p_1\leftrightarrow p_2)\bigg]\nonumber,
\end{align}
where \( \psi(x;p)\) stands for the Volkov solution of Dirac's equation with a classical external electromagnetic (laser) field \( A _\mu ( \phi) =a_0 \frac{m}{ \abs{ e}} \, \varepsilon _\mu f( \phi)  \), \( \overline \psi \) its adjoint, and \( D  _{ \mu \nu}  \) is the photon propagator. The \( p_1\leftrightarrow p_2 \) term ensures the antisymmetrization of two identical fermions (mass \( m \), charge \( \abs{ e} \)) in the final state. The laser field \( A _\mu \) and its polarisation four-vector \( \varepsilon _\mu  \) and phase \( \phi=k\cdot x \) is specialized further on below. The momenta \( p, p_{1,2,3} \) and \( k \) are four-vectors as well, and \( \gamma ^\mu  \) stands for Dirac's gamma matrices. Transforming the photon propagator into momentum space, \(D  _{ \mu \nu} (x,y)=\int d^4k'/ \left( 2 \pi\right)^4\, e^{-ik\cdot(x-y)} D_{ \mu \nu}(k')\) and employing the Feynman gauge, \( D_{ \mu \nu} (k') = \frac{g  _{ \mu \nu} }{k^{'2}+i \epsilon} \), and a suitable splitting of phase factors of the Volkov solution, e.g. \( \psi(x;p)=  \left( 1 + e\frac{\slashed k \slashed A}{2p\cdot k}\right)u_p\, e^{-i \hat S_p(k\cdot x)}\, e^{-ip\cdot x} \), one can cast the above matrix element in the form
\begin{align}\label{gl:matElement}
S_{fi}& = \left( 2 \pi\right)^2e^2\int dr \int ds \,\bigg[ \left(  \overline u(p_1) \Gamma ^\mu (r;C) u(p)\right) \frac{g  _{ \mu \nu} }{k'^2 + i \epsilon}\left(  \overline u(p_2) \Gamma ^\mu (s;BW) v(p_3)\right)
\nonumber\\ &\times\delta^{(4)}(p_1 + p_2 + p_3 - p -(r+s)k) - (p_1\leftrightarrow p_2)\bigg]
\end{align}
upon Fourier transform of the vertex function \( \left( 1+ \overline \Omega_{q_2}(k\!\cdot\!x)\right) \gamma^\mu  \left(1+ \Omega_{q_1}( k\!\cdot\!x)\right)\\\times e^{-i( \hat S_{q_1}(k\cdot x) - \hat S_{q_2}(k\cdot x))} = \int \frac{dr}{2 \pi} \Gamma ^\mu (r;q_1,q_2)e^{-ir\,k\cdot x} \). 
A key for that is \( \Omega_{p}( \phi=k\!\cdot\! x) = e\frac{\slashed k \slashed A}{2p\cdot k} \) as well as \( \hat S(\phi\!=\!k\!\cdot\! x;p)= \frac{1}{2k\cdot p}\int_{0}^{ k\cdot x} d \phi' \left( 2e\, p\!\cdot\! A(\phi') - e^2A^2(\phi')\right) \) as the nonlinear Volkov phase part. 
The quantit\new{ies} \( u(p) \) \new{and \( v(p) \) are} a free-field Dirac bispinor\new{s}, with spin \new{indices} suppressed for brevity. In intermediate steps, one meets the local energy-momentum balance \( p_1- p + k' - sk = 0 \) and \( p_2 + p_3 - k' - rk = 0 \), which combine to the overall conservation in \( \delta^{(4)} \).\ The corresponding representation of the matrix element in momentum space is exhibited in figure \ref{fig:feynDiagFull-b}-left, where we exposed the interaction with \( s \) and \( r \) laser photons marked by the crosses. In figure \ref{fig:feynDiagFull-b}-right, we suppress these explicit representations of the laser background field. Note that the momentum space diagrammatic\new{s} differs from the notation in \cite{Mitter:1975,Meuren:2013}.  We introduce the short-hand notations \( C \) and \( BW \) to mean momentum dependences on \( (p_1,-p) \) and \( (p_2,p_3) \), respectively. For the currents \( \Delta ^\mu (r,C) =  \overline u(p_1) \Gamma ^\mu u (p) \) and  \( \Delta ^\mu (s,BW) =  \overline u(p_2) \Gamma ^\mu v (p_3) \) in \eqref{gl:matElement}, we note the decomposition, emerging from inserting the Volkov solution, \( \Gamma ^\mu (r,*) = \Gamma_0 ^\mu (r,*) + \Gamma_1 ^\mu (r,*) + \Gamma_2 ^\mu (r,*) \) with \( * \) meaning \( C \) or \( BW \) and, by using a generic momentum pair \( \left( q_1,q_2\right) \) for them,
\begin{subequations}
\begin{align}
\Gamma_0 ^\mu(r,q_1,q_2)  &= \int_{-\infty}^{\infty} d \phi  e^ { i r \phi}e^{-i( \hat S_{q_1}(k\cdot x) - \hat S_{q_2}(k\cdot x))}\gamma^\mu,\\
\Gamma_1 ^\mu(r,q_1,q_2)  &= \int_{-\infty}^{\infty} d \phi e^ { i r \phi}e^{-i( \hat S_{q_1}(k\cdot x) - \hat S_{q_2}(k\cdot x))}\left[ \overline \Omega_{q_2}( \phi) \gamma^\mu  + \gamma^\mu \Omega_{q_1}(\phi)\right] ,\\
\Gamma_2 ^\mu(r,q_1,q_2)  &= \int_{-\infty}^{\infty} d \phi e^ { i r \phi}e^{-i( \hat S_{q_1}(k\cdot x) - \hat S_{q_2}(k\cdot x))} \left[ \overline \Omega_{q_2}( \phi) \gamma^\mu \Omega_{q_1}(\phi)\right] ,
\end{align}
\end{subequations}
which can be combined to arrive at the notation such as those of \cite{Ilderton:2011,King:2018}:
\begin{align}\label{eq:vertexFunc}
\Delta ^\mu (r,*) = \sum_{l=0}^2 B_l(r,*)J_l ^\mu (*),
\end{align}
with the currents
\begin{subequations}\label{eq:test}
\begin{align}
 J_0 ^\mu (q_1,q_2) &= \overline u(q_1) \gamma^\mu \Psi(q_2),\\
 J_{1} ^\mu (q_1,q_2) &= \frac{m a_0}{2} \overline u(q_1) \left[ \frac{\slashed \varepsilon \slashed k}{kq_1} \gamma^\mu  - \gamma^\mu \frac{ \slashed k\slashed \varepsilon}{kq_2}\right] \Psi(q_2),\\
J_2 ^\mu (q_1,q_2)&= -\frac{m^2 a_0^2}{2} \frac{k ^\mu }{(kq_1)(kq_2)}  \overline u(q_1) \slashed k \, \Psi(q_2),
\end{align}
\end{subequations}
where \( \Psi(q) = u(q) \) for \( C \) and \( \Psi(q) = v(q) \) for \( BW \), respectively. The phase integrals in equation \eqref{eq:vertexFunc} read 
\begin{align}\label{gl:phaseInt}
B_l(s,*) = \int_{-\infty}^{\infty} d \phi\, f^l( \phi) \exp \left\{i \sum_{n=0}^2 \alpha_n(s,*) \int_{0}^{ \phi} d \phi' f^n( \phi')\right\},
\end{align}
with
\begin{subequations}
\begin{align}
\alpha_0(s;q_1,q_2) &= s,\\
\alpha_{1}(s;q_1,q_2) &= ma_0 \left( \frac{q_1 \epsilon}{k q_1} - \frac{q_2 \epsilon}{kq_2}\right),\\
\alpha_2(s;q_1,q_2) &= \frac{m^2 a_0^2}{2} \left( \frac{1}{kq_1} + \frac{1}{kq_2}\right).
\end{align}
\end{subequations}
Note that in \eqref{gl:phaseInt}, \( l \) is an index (label) on the l.h.s, while it is a power on the r.h.s., as \( n \) too. An important step is the isolation of the divergent part in \( \Gamma_0 ^\mu  \) or \( B_0 \). We note \( \lim_{A\to 0}  \Gamma ^\mu(s,*)  = 2 \pi \delta(s) \gamma^\mu \) and regularise \( \Gamma ^\mu _0 \) by inserting a damping factor \( e^{-\epsilon \abs{ \phi}} \) and performing the limit \( \epsilon\to 0 \), similar to the method in \cite{Boca:2009}, which results in \( \Gamma ^\mu_0 (s,q_1,q_2) =  \mathcal G \pi\, \delta(s)\gamma^\mu  + \overset{12}{ \Gamma ^\mu_0 },\text{ with }\overset{12}{ \Gamma ^\mu_0 }   = - \mathcal P \frac{1}{s}\int_{-\infty}^{\infty} d \phi\, \gamma^\mu e^{is \phi} e^{-i( \hat S_{q_1}(\phi) - \hat S_{q_2}(\phi))}\frac{ \partial}{\partial \phi}( \hat S_{q_1}(\phi) - \hat S_{q_2}(\phi))
 \) or
\begin{align}\label{gl:B0reg}
B_0(s,*) &=  \mathcal G \pi \delta(s) - \mathcal P \left[ \frac{1}{s} \sum_{j=1}^2 \alpha_j(s,*) B_j(s,*)\right],\\
\mathcal G &=  \exp \left( \sum_{i=1}^2 \alpha_i \int_0^{+ \infty} f^i( \phi) \,d \phi\right) +  \exp \left( \sum_{i=1}^2 \alpha_i \int_0^{- \infty} f^i( \phi) \,d \phi\right)
\end{align}
where \( \mathcal P \) means the principal value in the variable \( s \). One can exploit in reading \eqref{eq:vertexFunc} the crossing symmetry \( \Delta ^\mu (r,BW=(p_2,p_3)) = \Delta ^\mu (r \to s,BW\to C=(p_1,-p)) \). 
Employing \eqref{gl:B0reg} with the short-hand notation \( \tilde B_0 = - \mathcal P \left[ \frac{1}{s} \sum_{j=1}^2 \alpha_j B_j(s,*) \right]\) and \( \tilde B_{1,2}=B_{1,2} \) we arrive at 
\begin{align}
g  _{ \mu \nu} \Delta ^\mu (r;C) \Delta^\nu (s;BW)  &= \pi^2 \mathcal G^2\delta(r) \delta(s) M_0 \nonumber\\&+ \pi \mathcal G \delta(r)M_{11}(s) + \pi \mathcal G \delta(s)M_{12}(r) \nonumber\\&+ M_{2}(r,s),\label{gl:deltadelta}
\end{align}
where
\begin{subequations}
\begin{align}
M_0 &= g  _{ \mu \nu}J_0^{\mu}(C) J_{0}^\nu (BW),\\
M_{11}(s) &= g  _{ \mu \nu}J_0^{ \mu}(C) \left( \sum_{l=0}^2\tilde B_{l}(BW)J_{l} ^\nu ({BW })\right),\label{eq:M11}\\
M_{12}(r) &= g  _{ \mu \nu}J_0^{\mu}(BW) \left(  \sum_{l=0}^2\tilde B_{l}({C})J_{l} ^\nu ({C })\right),\label{eq:M12}\\
M_{2}(r,s)&=g  _{ \mu \nu} \left( \sum_{l=0}^2\tilde B_{l}({s;BW})J_{l} ^\mu ({BW })\right)\left( \sum_{l=0}^2\tilde B_{l}(s;C)J_{l} ^\nu (C)\right).
\end{align}
These four expressions correspond to the momentum space diagrams exhibited in figure \ref{fig:feynDiagFull-c}.
\end{subequations}
\section{Weak-field expansion}\label{sec:wf}
\begin{figure}[t]
\centering
\includegraphics[width=0.7\textwidth]{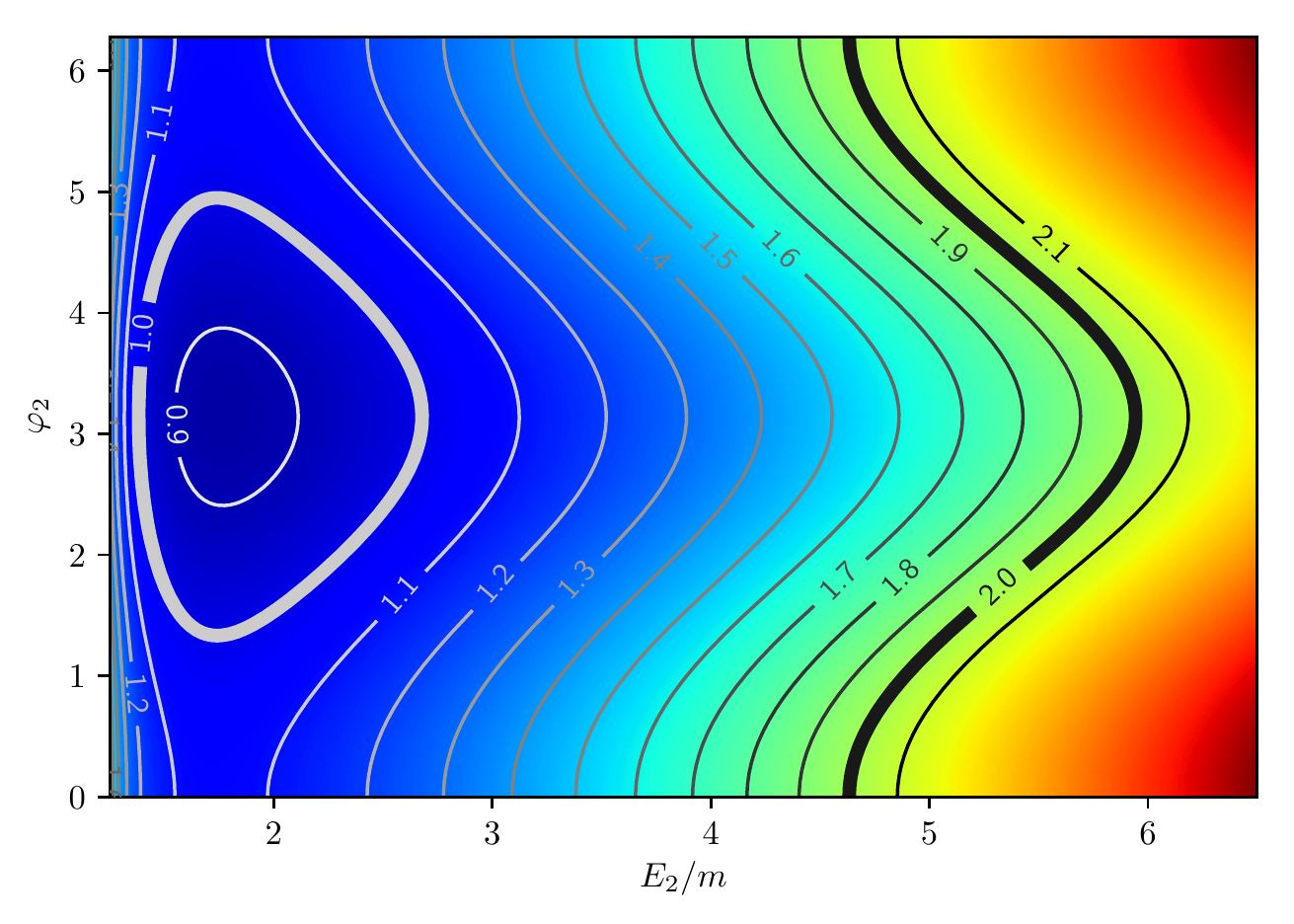}
\caption{ Contour plot \( s(E_2, \phi_2) \) for \( E_3 = 1.76\, m \), \( \cos\theta_{2,3} = 0.965 \) and  \( \varphi_3 = 0 \). The initial electron is at rest in this frame, and laser frequency amounts to \(\new{\omega=}\,  k_0 = 5.12\, m, \) i.e.\ the center of momentum energy is \( \sqrt{ \left( k+p\right)^2} = 3.353\,m \).}
\label{fig:SvsEphi}
\end{figure}

With the argumentation given in the introduction we now attempt an expansion in powers of \( a_0 \).
We note \( J ^\mu _l \propto a_0^l \) and \( \alpha_l \propto a_0^l \); again, \( l \) is a label (power) on the l.h.s.\ (r.h.s.).
The leading-order terms of the phase integrals \eqref{gl:phaseInt} thus become
\begin{subequations}
\begin{align}
\tilde B_0(s) &=  - \mathcal P \left[ \frac{a_0}{s} \tilde{\alpha}_1 \int_{-\infty}^{\infty} d \phi f (\phi)e^{is \phi} \right]+ \mathcal O(a_0^2),\\
\tilde B_j(s) &= \int _{-\infty}^{\infty}d \phi f^j( \phi)e^{is \phi} \left[ 1 +  i a_0\tilde{ \alpha}_1\int_{0}^{ \phi} d \phi' f( \phi')  \right] + \mathcal O(a_0^2),\\
\mathcal G &= 2 + ia_0 \tilde\alpha_1 \lim_{ \eta\to \infty}\left( \int_{0}^{\eta} f( \phi)\, d\phi +\int_{0}^{-\eta} f( \phi)\, d\phi \right)+\mathcal O(a_0^2).
\end{align}
\end{subequations}
Denoting the Fourier transform of \( f( \phi) \) by
\begin{align}\label{eq:generalF}
F(s) = \int_{-\infty}^{\infty}d \phi\, f( \phi) \exp\{is \phi\}
\end{align}
we recognize that 
\begin{subequations}
\begin{align}
M_0 &\propto a_0^0,\label{eq:M0weak}\\
M_{11}(s) &= a_0 J_0^{\mu}(C) \bigg( -\mathcal P \left[ \frac{ \tilde{\alpha}_1(BW)}{s} F(s)\right]J_{0 \mu}({BW}) + \tilde J_{1 \mu}({BW })F(s)\bigg)+ \mathcal O(a_0^2),\label{eq:M11weak}\\
M_{12}(r) &= a_0 J_0^{\mu}(BW) \bigg( -\mathcal P\left[ \frac{\tilde{\alpha}_1({C}) }{r} F(r) \right]J_{0 \mu}({C}) +  \tilde J_{1 \mu}({C })F(r)\bigg)+\mathcal O(a_0^2),\label{eq:M12weak}\\
M_{2}(r,s)& \propto \mathcal O(a_0^3)\label{gl:M2weak}.
\end{align}
\end{subequations}
The two delta distributions in the \( M_0 \) term in \eqref{gl:deltadelta} enforce for the overall momentum conservation in \eqref{gl:matElement} a factor \( \delta^{(4)}(p_1 + p_2 + p_3 - p)   \) implying a zero contribution of the \( M_0 \) term \new{\eqref{eq:M0weak}}. \new{In the spirit of the \( a_0 \) series expansion we neglect \eqref{gl:M2weak} at all. (This term would give rise to on/off-shell contributions which require some care.)} The remaining leading order terms in \new{\eqref{eq:M11weak} and \eqref{eq:M12weak}} generate the contributions
\begin{subequations}
\begin{align}
S_s&= 4 \mathcal G\pi^3e^2 \int_{-\infty}^{\infty} ds\frac{ M_{11}(s)}{(p-p_1)^2 + i \epsilon} \delta^{(4)}(p_1 + p_2 + p_3 -p -sk),\\
S_r &= 4 \mathcal G\pi^3e^2 \int_{-\infty}^{\infty} dr \, \frac{ M_{12}(r)}{(p_2+p_3)^2 + i \epsilon}\delta^{(4)}(p_1 + p_2 + p_3 -p -rk)
\end{align}
\end{subequations}
plus the corresponding exchange terms upon \( p_1 \leftrightarrow p_2 \). Introducing light-front coordinates\footnote{\new{We use the definition \( q^{\pm} = \frac{1}{2} \left( q^0 \pm q^3\right) \) and \( q^{\perp}=(q^1, q^2) \) for the light-front coordinates of a four-vector \( q ^\mu = (q^0, q^1, q^2, q^3) \).}} \new{and the light-front delta distribution \( \delta^{ \mathrm{lf}}(p) = \delta ( p^-) \delta^{(2)}  ( p^\perp)\) as well as choosing \( k^+ \) as the non-zero component of the laser four-momentum \( k ^\mu  \)} yields 
\begin{subequations}
\begin{align}
S_s &= \frac{2\mathcal G\pi^3e^2}{k^+}\delta^{\mathrm{lf}}(p_1 + p_2 + p_3 -p)\frac{ M_{11}(s)}{(p-p_1)^2 + i \epsilon},\\
S_r &=\frac{ 2\mathcal G\pi^3e^2}{k^+}\delta^{\mathrm{lf}}(p_1 + p_2 + p_3 -p )\frac{ M_{12}(r)}{(p_2+p_3)^2 + i \epsilon},
\end{align}
\end{subequations}
where \( M_{11} \) refers to \eqref{eq:M11} and \( M_{12} \) to \eqref{eq:M12} and one has to use in both cases
\begin{align}\label{eq:defPhotoNum}
r = s = \frac{p^+ - p_1^+ - p_2^+ - p_3^+}{k^+} \neq 0,
\end{align}
and the principal value can be dropped due to the last inequality. For given entry channel parameters, equation \eqref{eq:defPhotoNum} implies the dependence \( s( E_{2,3}, \cos\theta_{2,3}, \varphi_{2,3}) \) with spherical coordinates \( E_i, \cos \theta_i, \varphi_i \) for the particles \( 1, 2 \) and \( 3 \) (cf.\ figure \ref{fig:feynDiagFull-a}) in the exit channel. Fixing \( E_3, \cos\theta_{2,3}  \) and \( \varphi_3 \) yields the contour plot \( s(E_2, \varphi_2) \). An example is exhibited in figure \ref{fig:SvsEphi}. The locus of the \new{apparent} one- (two-) photon contribution with \( s=1 \) (\( =2 \)) is highlighted by fat curves.
\\
The final result is the leading-order matrix element
\begin{align}\label{eq:leadOrdMat}
S_{fi} &=\frac{ 2\mathcal G\pi^3e^2}{k^+} a_0\left[ \frac{M(s;BW)}{(p-p_1)^2 + i \epsilon}+\frac{M(s;C)}{(p_2+p_3)^2 + i \epsilon}\right] F(s)\,\delta^{\mathrm{lf}}(p_1 + p_2 + p_3 -p ) 
\end{align}
with 
\begin{subequations}
\begin{align}
M(s;BW) &= g  _{ \mu \nu} J_0^{\mu}(C) \bigg(\frac{ \widetilde{\alpha}_1(BW)}{s} J_{0} ^\nu (BW)+ \widetilde J_{1} ^\nu (BW)\bigg),\label{eg:MBW}\\
M(s;C) &= g  _{ \mu \nu} J_0^{ \mu}(BW) \bigg(\frac{\widetilde{\alpha}_1(C) }{s} J_{0 } ^\nu (C) +  \widetilde J_{1 } ^\nu (C)\bigg),\label{eg:MC}
\end{align}
\end{subequations}
where the tildes indicate here that the factor \( a_0 \) is scaled out. These structures are suggestive: \( M(BW) \) may be read as the coupling of the free Compton current \( J_0 ^\mu (C) \) to the modified Breit-Wheeler current (in parenthesis of \eqref{eg:MBW}) and a free Breit-Wheeler current \( J_0 ^\mu (BW) \) to a modified Compton current (in parenthesis of \eqref{eg:MC}), both depicted in the middle panels of figure \ref{fig:feynDiagFull-c}. The interaction with the external field is encoded in the modified currents. For practical purposes we replace the modified currents by the proper Volkov currents \( \Delta ^\mu (C) \) and \( \Delta ^\mu (BW) \) when evaluating numerically  the matrix elements for \( a_0 \ll 1 \). The differential probability is
\begin{align}\label{eq:diffProb}
dw &= \left(  \frac{  2\mathcal G\pi^3e^2}{k^+}\right)^2\abs{ \mathcal M}^2\new{(2 \pi)^3}\delta^{\mathrm{lf}}(p_1 + p_2+ p_3 -p )  \abs{ F(s)}^2 V_{\mathrm{lf} } \, d \Pi_{3}
\end{align}
with three-body phase space element
\begin{align}
d \Pi_3 = \frac{\Theta(p_1^-) \Theta(p_2^-) \Theta(p_3^-)}{(2 \pi)^9} \frac{dp_1^-\, d^2 p_1^{\perp}}{2 p_1^-} \frac{dp_2^-\, d^2 p_2^{\perp}}{2 p_2^-} \frac{dp_3^-\, d^2 p_3^{\perp}}{2 p_3^-},
\end{align}
\neww{the Heaviside step-function \( \Theta \)} \new{and the differential cross section is 
\begin{align}
d \sigma = \frac{dw}{NV_{\mathrm{lf} }}\label{eq:WWQ}
\end{align}
with the light-front volume \( V_{\mathrm{lf} } \) and the normalisation factor \( N = \frac{a_0^2 m^2}{2 e^2}\int_{-\infty}^{\infty} g^2( \phi) \,d \phi  \) (cf. \cite{Seipt:2011}).}
The matrix element \( \mathcal M \) is given by \eqref{eq:leadOrdMat} but without the pre-factor:
\begin{align}
\mathcal M =  \frac{M(s;BW)}{(p-p_1)^2 + i \epsilon}+\frac{M(s;C)}{(p_2+p_3)^2 + i \epsilon}.
\end{align}
We emphasize  that the weak-field limit \eqref{eq:leadOrdMat} in \eqref{eq:diffProb} corresponds to the standard perturbative tree level QED diagrams depicted in the figure \ref{fig:FeynDiagPert}, supposed
\begin{align}\label{eq:generalFlimit}
\abs{ F(s)}^2 \to \delta(s-1) + \delta(s+1).
\end{align}
\( s=1 \) selects then the admissible kinematics, and the phase space in \eqref{eq:diffProb} becomes five-dimensional. The decomposition \eqref{gl:B0reg} is essential for catching the proper weak-field perturbative limit. This is obvious when considering M\o ller or Bhabha scattering in an ambient background field as the cross channel\new{s} of \new{the} trident process: For \( A\to 0 \) the standard perturbative QED must be recovered.
\section{The case of a \( \mathbf{\cos^2} \) envelope}\label{sec:cos2}
We now specialise the linearly polarised external e.m. field \( A ^\mu  \) with \( \varepsilon ^\mu = \left( 0,1,0,0\right) \) to a \( \cos^2 \) envelope times the oscillating part yielding
\begin{align}\label{eq:cosSQfield}
f( \phi) =  \left[ \cos^2 \left( \frac{ \pi \phi}{2 \Delta \phi}\right)\, \new{ \bigsqcap \left( \frac{\phi}{2\Delta \phi}\right)}\right] \, \cos( \phi + \phi_{ \mathrm{cep}}),
\end{align}
where \new{\( \bigsqcap\left( \frac{\phi}{2\Delta \phi}\right) \)} is a box profile of width \( 2 \Delta \phi \) centred at \( \phi=0 \). We leave a discussion of the carrier envelope phase \( \phi_{ \mathrm{cep}} \) for separate work, i.e. put \( \phi_{ \mathrm{cep}} =0\). Then \eqref{eq:generalF} can be integrated analytically with the result
\begin{align}
\frac{ 1}{\Delta \phi}F(s, \Delta \phi)&= \frac{ \pi^2}{2} \left[ \frac{\operatorname{sinc}\left( \Delta \phi(s-1)\right)}{ \pi^2 - \Delta \phi^2(s-1)} +  \frac{\operatorname{sinc}\left( \Delta \phi(s+1)\right)}{ \pi^2 - \Delta \phi^2(s+1)}\right],\label{eq:cosSQF}
\end{align}
where \( \operatorname{sinc}(x) = \frac{\sin(x)}{x} \) for \( x\neq 0 \) and \( \operatorname{sinc}(0)=1 \) is the cardinal sine function. With a proper normalisation, \eqref{eq:diffProb} is proportional to \( \frac{ \abs{ F(s)}^2}{ \Delta \phi} \). \( F(s) \) is a real function  due to the even symmetry of the special field \eqref{eq:cosSQfield}, but in general it aquires also an imaginary part. We therefore keep the notation \( \abs{ F(s)}^2 \).
\begin{figure}[t]
\centering
\includegraphics[width=0.7\textwidth]{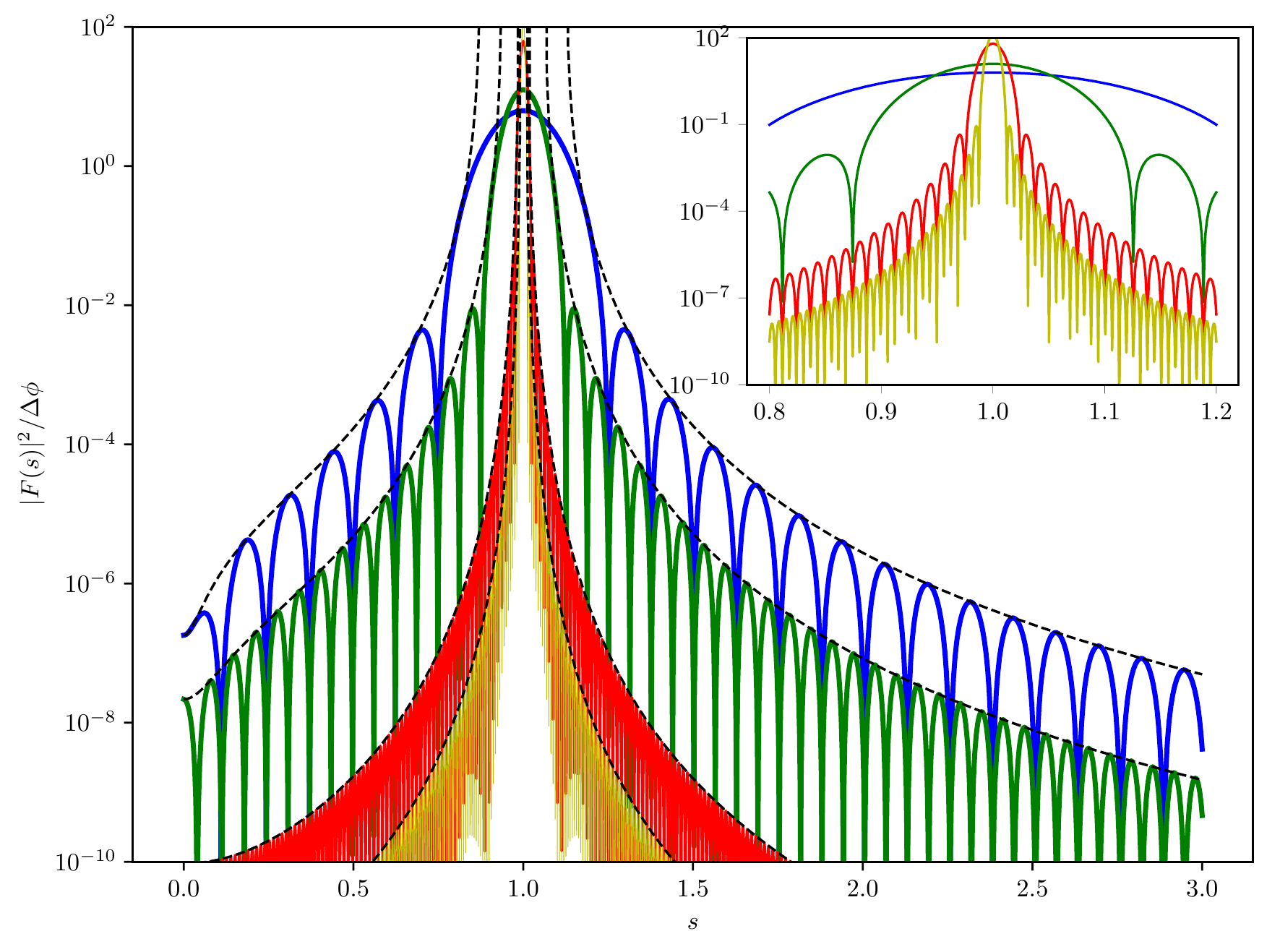}
\caption{The quantity \( \frac{ \abs{ F(s)}^2}{ \Delta \phi} \) as a function of \( s \) for several values of \(\Delta \phi \) (blue: \( \Delta \phi = 25 \), green: \( \Delta \phi =50 \), red: \( \Delta \phi =250 \), yellow: \( \Delta \phi = 500 \)). Dashed curves exhibit the envelopes according to equation \eqref{eq:cosSQF}. The inset zooms into the region \( s\approx 1 \). Note the symmetry property \( \abs{F(s)}^2 = \abs{F(-s)}^2 \).}
\label{fig:FvsS}
\end{figure}
Given the properties of the sinc functions entering \eqref{eq:cosSQF} we find
\begin{align}
\lim_{ \Delta \phi\to \infty}\frac{\abs{F(s)}^2}{ \Delta \phi} = \frac{1}{4} \left( \delta(s+1) + \delta(s-1)\right),
\end{align}
thus making \eqref{eq:generalFlimit} explicit. Since \( \frac{\abs{F(s)}^2}{ \Delta \phi} >0 \) for \( s \neq 1 \), in particular for \( s>1 \), at finite values of \( \Delta \phi \), we see that this signals \new{bandwidth} effects, despite \( a_0\ll 1 \). Such effects have been observed in \cite{Nousch:2012} for the Breit-Wheeler pair production below the threshold and the Compton process as well \cite{Titov:2013,Titov:2012}. Since \( s \) is a continuous variable one must not identify it with a ``photon number''; instead, \( s \) could be interpreted as fraction of energy or momentum in units of \( \omega = \abs{ \vec k} \) participating in creating a final state different from the initial state (see \cite{Ritus:1985} for discussions of that issue). \new{Even more, \( s>1 \) does not mean proper multiphoton effects due to our restriction on leading order in \( a_0 \), rather one could speak an ``apparent multiphoton effects'' caused by finite bandwidth of the pulse.}
\begin{figure}[t]
\includegraphics[width=\textwidth]{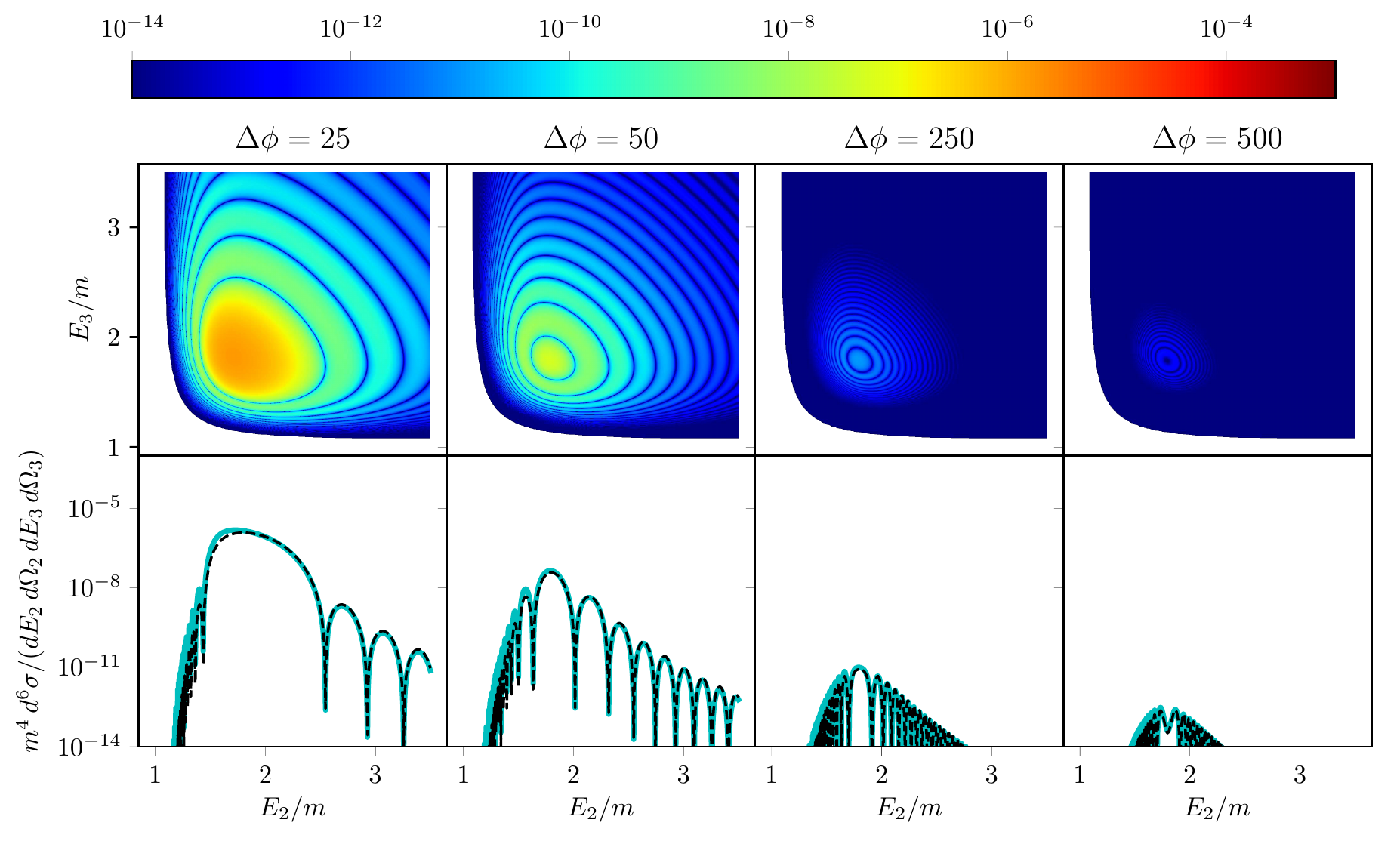}
\caption{The differential cross section \( d^6 \sigma/ \left( dE_2\,d^2\Omega_2\,dE_3\,d^2\Omega_3\right)\) \new{from equation \eqref{eq:WWQ} with \eqref{eg:MBW} and \eqref{eg:MC}} for an intensity parameter \( a_0 = 10^{-4} \) with \( d^2\Omega_{2,3} = d\cos \theta_{2,3} d \varphi_{2,3}\) for \( \cos\theta_{2,3} = 0.95 \), \( \varphi_{2} = \pi/2 \), \( \varphi_3 = 0 \) over the \(  E_2 - E_3 \) plane (top) and as a function of \( E_2 \) for \( E_3 = 1.76 m \) (bottom, solid \new{cyan} curves). \new{The initial electron is at rest and the laser frequency amounts to \(\omega = k^0 = 5.12\,m \) in this frame.} The pulse length parameters are \( \Delta \phi = 25,\dots,500 \) as indicated. In the bottom panels, the function \( \abs{ F(s, \Delta \phi)}^2/ \Delta \phi \) is exhibited by dashed black curves, scaled up by a common factor of \( 5.2\times 10^{-7} \).} 
\label{fig:CSforbidden}
\end{figure}
The dependence of \( \frac{\abs{F(s)}^2}{ \Delta \phi}  \) on \( s \) for various values of \( \Delta\phi \) is displayed in figure \ref{fig:FvsS}. The function is symmetric, \( \abs{F(s)}^2 = \abs{F(-s)}^2 \), with main maxima at \( s=\pm 1 \) and the envelopes are
\begin{align}
\frac{\overset{ \mathrm{env}}{\abs{F(s)}^2}}{\Delta \phi} = \frac{1}{ \Delta \phi}\abs{ \int_{-\infty}^{\infty}d \phi\, 2 \Theta( \phi) f( \phi)e^{is \phi}}^2.
\end{align}
\new{We use the absolute value of the analytic signal of \( F(s) \), i.e.\ the Fourier transform of the amplitude function \eqref{eq:cosSQfield} constrained to the positive half-line \cite{Cohen:1995}.}
The first side maxima are located between \( 1\pm 2\frac{ \pi}{ \Delta \phi} \) and  \( 1\pm 3\frac{ \pi}{ \Delta \phi} \) and their heights are \( 7\times 10^{-4} \) of the respective main maximum\new{, meaning that their contribution is not negligible, in particular for a kinematical situation where \( s>1\) (cf.\ figure \ref{fig:SvsEphi})}\\
From the definition of \( s \) in equation \eqref{eq:defPhotoNum}, we note (i) the dependence \( s(E_2,E_3, \theta_{2,3}, \varphi_{2,3}) \) and (ii) the U shape of \( s(E_2) \) when keeping constant the other variables of the momenta in polar coordinates. Denoting the minimum of \( s(E_2) \) by \( s_{ \mathrm{min}} \), then for kinematical situations, where \( s_{ \mathrm{min}}>1 \) the side peaks of \( \frac{\abs{F(s)}^2}{ \Delta \phi} \) with spacing of about \( \frac{ \pi}{ \Delta \phi} \) become relevant. In fact, \( \frac{\abs{F(s)}^2}{ \Delta \phi} \) carries much of the energy dependence of the differential cross section.
As an example, we exhibit in figure \ref{fig:CSforbidden} (bottom) the differential cross section \( d^6 \sigma/ \left( dE_2\,d\cos \theta_2\, d \varphi_2\,dE_3\,d\cos \theta_3\, d \varphi_3\right) \) \new{(multiplied by \( m^4 \) to make it dimensionless)} as a function of \( E_2 \) \new{and compare it with} \( \frac{\abs{F(s(E_2), \Delta \phi)}^2}{ \Delta \phi} \) \new{scaled by a factor \( 5.2\times 10^{-7} \)}. Note the near-perfect match.\footnote{\new{Supposed the match of the cross section \eqref{eq:WWQ} and \(  \frac{\abs{F(s, \Delta \phi)}^2}{ \Delta \phi} \) scaled by one common factor continues over a wide region in phase space one could envisage a greatly simplified numerical procedure by evaluation (or even estimating) \eqref{eq:WWQ} only a few times and continue it with \(  \frac{\abs{F(s, \Delta \phi)}^2}{ \Delta \phi} \).}}
\begin{figure}[t]
\includegraphics[width=\textwidth]{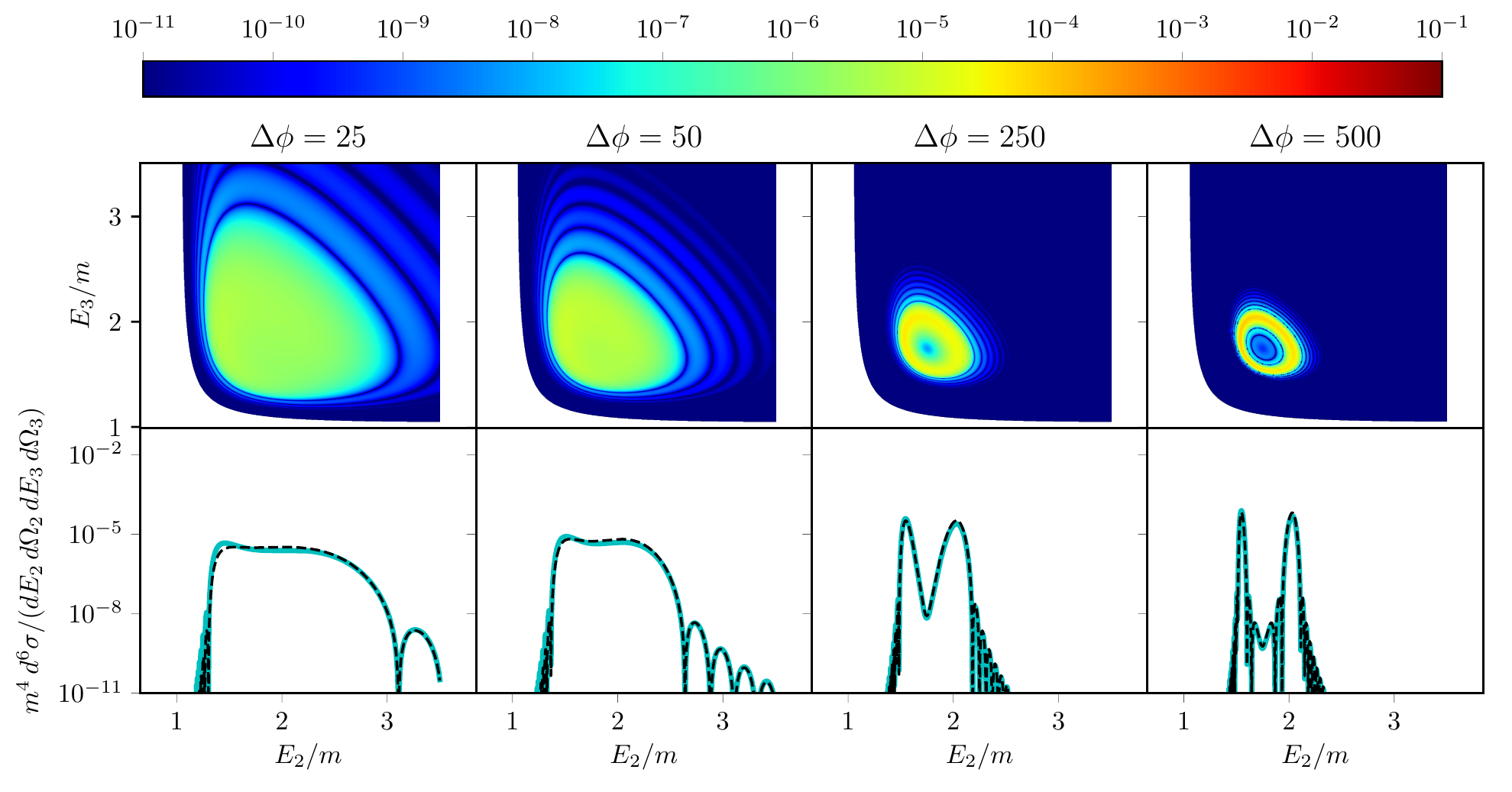}
\caption{As figure \ref{fig:CSforbidden} but for \( \cos\theta_{2,3} = 0.965\) and the same \new{scaling} factor \new{of \(  \frac{\abs{F(s, \Delta \phi)}^2}{ \Delta \phi} \)}. Again a near perfect agreement of the solid and dashed curve\new{s} is achieved.}
\label{fig:CSnewKin}
\end{figure}
The selected kinematics implies \( s_{ \mathrm{min}} \approx 1.1\), i.e.\ in the limit \( \Delta \phi \to \infty \), this setting would be kinematically forbidden, but \new{bandwidth} effects for finite values of \( \Delta \phi \) enable the selected kinematics. In fact, increasing \( \Delta \phi \) causes \( (i) \) a rapid dropping of the differential cross section and \( (ii) \) make the oscillatory pattern more dense. At the heart of the behavior is essentially the quantity   \( \frac{\abs{F(s, \Delta \phi)}^2}{ \Delta \phi} \) of figure \ref{fig:FvsS}; the additional \( E_2 \) dependence is fairly smooth. To translate \(  \frac{\abs{F(s, \Delta \phi)}^2}{ \Delta \phi} \) of figure \ref{fig:FvsS} into the bottom panels of figure \ref{fig:CSforbidden} one needs explicitly \( s(E_2) \) which can be inferred from figure \ref{fig:SvsEphi}. \\
Displaying the differential cross section as contour plot over the \( E_2 - E_3 \) plane (see figure \ref{fig:CSforbidden}-top) and keeping \( \theta_{2,3} \) and \( \varphi_{2,3} \) fixed as above, one observes a pronounced fringe pattern which become denser with increasing values of \( \Delta \phi \). The fringe pattern occupies a finite region in the \( E_2-E_3 \) plane. At the origin of the fringe pattern is again the function \( \frac{\abs{F(s(E_2), \Delta \phi)}^2}{ \Delta \phi} \). The fringe pattern make numerical integrations towards total cross section fairly challenging.\\
The relevance of the function \( \frac{ \abs{ F}^2}{ \Delta \phi} \) continues of course for kinematical situations which are not forbidden in the limit \( \Delta \phi \to \infty\). 
Examples are exhibited in figure \ref{fig:CSnewKin}. For the selected kinematic situation, \( s_{min} = 0.976 \), meaning that with increasing values of \( \Delta \phi \) the differential cross section does not drop but gets concentrated at two values of \( E_2 \) (at given \( E_3 \)) which are allowed for \( s=1 \). In the limit \( \Delta \phi \to \infty \), two delta peaks arise when \( E_3 \) is appropriately fixed. They are the result of cutting the strength distribution over the \( E_2-E_3\) plane at \( E_3=\text{const.} \), i.e.\ there is a sharp ring given by the solution of \( E_3(E_2) \). The tilt of the U shaped function \( s(E_2) \) at fixed other parameters to the right makes the region \( s\approx 1 \) larger at large but finite values of \( \Delta \phi \). Correspondingly, the r.h.s\ peak structure is wider, as seen e.g.\ in the right bottom panel.
\begin{figure}[t]
\centering
\includegraphics[width=0.7\textwidth]{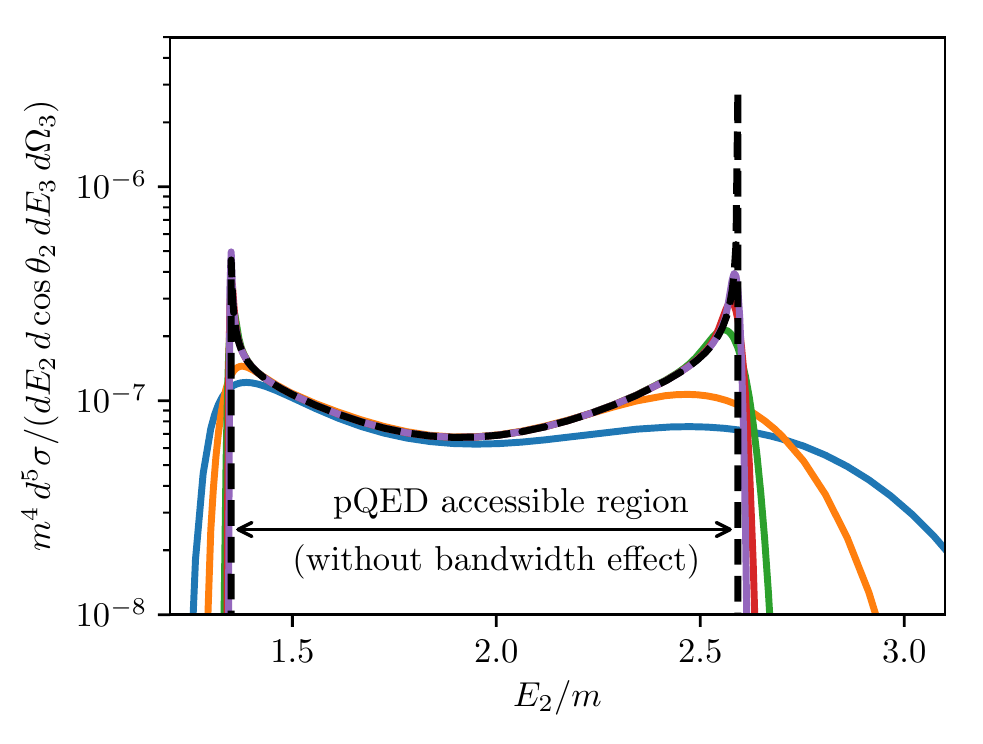}
\caption{The differential cross section \(  d^5 \sigma/ \left( dE_2\,d\cos \theta_2\,dE_3\,d^2\Omega_3\right) \) \new{from \eqref{eq:WWQ} with \eqref{eg:MBW} and \eqref{eg:MC}} as a function of \( E_2/m \) for \( E_3 = 2.0\,m \), \( \cos\theta_{2,3} = 0.965 \) and \( \varphi_3 = 0 \) for various values of \( \Delta \phi \) (solid curves, \( \Delta \phi = 25 \text{: blue}, 50 \text{: orange}, 250 \text{: green}, 500 \text{: red} \)) \new{at \( a_0 = 10^{-4} \)}. \new{The initial electron is at rest and the laser frequency amounts to \( \omega=k^0 = 5.12\,m \) in this frame.}\ The pQED result is depicted by the black dashed curve (we checked our \new{pQED-}software package by comparing with \cite{Haug:1975,Jarp:1973,Mork:1967} \new{and get confidence of our numerical evaluation and normalisation of \eqref{eq:WWQ} by the smooth approach towards the pQED result for large values of \( \Delta \phi \)}).}
\label{fig:pertLimit}
\end{figure}\\
To compare with the \new{standard} perturbative QED (pQED) result, based \new{o}n \new{the} evaluation of the \new{Feynman} diagrams \new{depicted} in figure \ref{fig:FeynDiagPert}, one has to perform the \( \varphi_2 \) integration. In fact, making \( \Delta \phi \) larger and larger, the differential cross section \( d^5 \sigma/ \left( dE_2\,d\cos \theta_2\,dE_3\,d^2\Omega_3\right)  \) \new{from equation \eqref{eq:WWQ}} approaches the pQED result \new{which is for a monochromatic photon beam}, see figure \ref{fig:pertLimit}. This figure illustrates how the pQED is approached for \( a_0\ll 1 \) and \( \Delta \phi\to \infty \). It also demonstrates that for \( \Delta \phi< 100 \) a significantly larger phase space beyond the perturbatively accessible \new{(one-photon)} region \new{(indicated by \( \longleftrightarrow \))} is occupied due to \new{bandwidth} effects \new{in laser pulses}.
\section{Summary}\label{sec:conc}
The length of laser pulses has a decisive impact on the pair production in the trident process. The rich phase space patterns, already found and analysed in some detail for non-linear one-vertex processes à la Compton and Breit-Wheeler, show up also in the two-vertex trident process. Even for weak laser fields, a region becomes accessible which would be kinematically forbidden in a strict perturbative, leading-order tree level QED approach. The key is the frequency distribution in a pulse which differs significantly from a monochromatic laser beam, which would mean an ``infinitely long laser pulse''. By resorting to a special pulse model \( \propto \cos^2 \left( \pi \phi/2 \Delta \phi\right)\cos( \phi) \), we quantify in a transparent manner the effect of the pulse duration \( \Delta \phi \) and identify the transition to monochromatic laser fields, \( \Delta \phi \to \infty \), for small values of the classical laser non-linearity parameter. To ensure the contact to a perturbative QED approach the proper decomposition of a phase factor is mandatory. To be specific, the first term in \eqref{gl:B0reg} is essential for the two-vertex process; for one-vertex processes it does not contribute. The effect of short laser pulses manifests itself in \new{bandwidth effects mimicking apparent} multiphoton contributions even for weak fields, where the intermediate photon is off-shell and and the process is of one-step nature. The rich pattern of the differential phase space distribution is traced back to the Fourier transform of the external e.m.\ field. In obvious further extensions of our approach, more general pulse envelope shapes should be studied, e.g.\ within the slowly varying envelope approximation. Going to larger values of the classical laser non-linearity parameter means checking whether an analog of the Fourier transform of the e.m.\ field can be isolated as crucial element of the phase space distribution of produced particles\footnote{\new{In fact, for \( a_0<0.01\) and large values of \( \Delta \phi \) we find numerical agreement of \eqref{eq:WWQ} and the pQED result.}}. The final state phase distribution is important for planing corresponding experimental designs.
\ack
The authors gratefully acknowledge the collaboration with A. Otto and very useful discussions with F. Mackenroth. We appreciate conversations with our colleges \mbox{B. King}, A. M. Fedotov, G. Torgrimsson, A. Di Piazza, A. Ilderton, T. Heinzl, A. Hartin, \mbox{A. I. Titov}, D. Seipt, M. Bussmann and T. Nousch.
The work is supported by R. Sauerbrey and \mbox{T. E. Cowan} w.r.t.\ the study of fundamental QED processes for HIBEF.
\section*{References}
\bibliography{paper2018IOP}
\bibliographystyle{unsrt}

\end{document}